\documentclass[12pt]{article}
\usepackage{amssymb}

\def\hybrid{\topmargin 0pt      \oddsidemargin 0pt
        \headheight 0pt \headsep 0pt
        \voffset=-0.5cm
        \hoffset=-0.25in
        \textwidth 6.75in
        \textheight 9.5in       
        \marginparwidth 0.0in
        \parskip 5pt plus 1pt   \jot = 1.5ex}
\catcode`\@=11
\def\marginnote#1{}

\newcount\hour
\newcount\minute
\newtoks\amorpm
\hour=\time\divide\hour by60 \minute=\time{\multiply\hour by60
\global\advance\minute by-\hour}
\edef\standardtime{{\ifnum\hour<12 \global\amorpm={am}%
        \else\global\amorpm={pm}\advance\hour by-12 \fi
        \ifnum\hour=0 \hour=12 \fi
        \number\hour:\ifnum\minute<10 0\fi\number\minute\the\amorpm}}
\edef\militarytime{\number\hour:\ifnum\minute<10 0\fi\number\minute}

\def\draftlabel#1{{\@bsphack\if@filesw {\let\thepage\relax
   \xdef\@gtempa{\write\@auxout{\string
      \newlabel{#1}{{\@currentlabel}{\thepage}}}}}\@gtempa
   \if@nobreak \ifvmode\nobreak\fi\fi\fi\@esphack}
        \gdef\@eqnlabel{#1}}
\def\@eqnlabel{}
\def\@vacuum{}
\def\draftmarginnote#1{\marginpar{\raggedright\scriptsize\tt#1}}
\def\draftlabel#1{{\@bsphack\if@filesw {\let\thepage\relax
   \xdef\@gtempa{\write\@auxout{\string
      \newlabel{#1}{{\@currentlabel}{\thepage}}}}}\@gtempa
   \if@nobreak \ifvmode\nobreak\fi\fi\fi\@esphack}
        \gdef\@eqnlabel{#1}}
\def\@eqnlabel{}
\def\@vacuum{}
\def\draftmarginnote#1{\marginpar{\raggedright\scriptsize\tt#1}}

\def\draft{\oddsidemargin -.5truein
        \def\@oddfoot{\sl preliminary draft \hfil
        \rm\thepage\hfil\sl\today\quad\militarytime}
        \let\@evenfoot\@oddfoot \overfullrule 3pt
        \let\label=\draftlabel
        \let\marginnote=\draftmarginnote
   \def\@eqnnum{(\theequation)\rlap{\kern\marginparsep\tt\@eqnlabel}%
\global\let\@eqnlabel\@vacuum}  }

\def\numberbysection{\@addtoreset{equation}{section}
        \def\theequation{\thesection.\arabic{equation}}}

\def\underline#1{\relax\ifmmode\@@underline#1\else
        $\@@underline{\hbox{#1}}$\relax\fi}

\def\titlepage{\@restonecolfalse\if@twocolumn\@restonecoltrue\onecolumn
     \else \newpage \fi \thispagestyle{empty}\c@page\z@
        \def\thefootnote{\fnsymbol{footnote}} }

\def\endtitlepage{\if@restonecol\twocolumn \else  \fi
        \def\thefootnote{\arabic{footnote}}
        \setcounter{footnote}{0}}  

\numberbysection \hybrid

\newcounter{mo}

\newcommand{\tr}{{\rm tr}}
\newcommand{\ti}[1]{\tilde{#1}}

\newcommand{\mL}{{\mathcal L}}
\newcommand{\mM}{{\mathcal M}}
\newcommand{\mF}{{\mathcal F}}

\newcommand{\mH}{{\mathcal H}}
\newcommand{\mU}{{\mathcal U}}
\newcommand{\mO}{{\mathcal O}}
\newcommand{\mV}{{\mathcal V}}

\newcommand{\vf}{\varphi}
\newcommand{\al}{\alpha}
\newcommand{\be}{\beta}
\newcommand{\ga}{\gamma}
\newcommand{\om}{\omega}
\newcommand{\vth}{\vartheta}

\newcommand{\Mat}{ {\rm Mat}(N,\mathbb C) }
\newcommand{\MatM}{ {\rm Mat}(M,\mathbb C) }
\newcommand{\MatNM}{ {\rm Mat}(NM,\mathbb C) }

\newcommand{\mC}{\mathbb C}
\newcommand{\mZ}{\mathbb Z}

\newcommand{\mS}{\mathcal S}

\newcommand{\h}{\hbar}

\newtheorem{predl}{Proposition}[section]

\def\beq{\begin{equation}}
\def\eq{\end{equation}}
\def\p{\partial}

\begin{document}

\setcounter{page}{1}


\begin{flushright}
 ITEP-TH-37/18\\
\end{flushright}
\vspace{0mm}

\begin{center}
\vspace{0mm}
{\LARGE{ Generalized model of interacting integrable tops}}
 \vspace{4mm}
\\
\vspace{15mm} {\large \ \ {A. Grekov}\,{\small $^{\diamondsuit\,
\flat\,\ddagger\,\natural}$}
 \ \ \ \ \ \ \ {I. Sechin}\,{\small $^{\diamondsuit\, \flat\, \ddagger}$}
 \ \ \ \ \ \ \ {A. Zotov}\,{\small $^{\diamondsuit\, \ddagger\, \S\,
\natural}$} }
 \vspace{10mm}

\vspace{1mm} $^\diamondsuit$ -- {\small{\rm
 Steklov Mathematical Institute of Russian Academy of Sciences,\\ Gubkina str. 8, Moscow,
119991,  Russia}}\\
 \vspace{1mm} $^\ddagger$ -- {\small{\rm 
 ITEP, B. Cheremushkinskaya str. 25,  Moscow, 117218, Russia}}\\
 \vspace{1mm} $^\flat$ -- 
 {\small{\rm Center for Advanced Studies,
 Skolkovo Institute of Science and Technology,\\
 Nobel str. 1, Moscow,  143026, Russia}}
\\
 \vspace{1mm}$^\S$ - {\small{\rm National Research University Higher School of Economics, \\
 Usacheva str. 6,  Moscow, 119048, Russia}}
 \\
  \vspace{1mm} $^\natural$ -- {\small{\rm Moscow Institute of Physics and Technology,\\ Inststitutskii per.  9, Dolgoprudny,
 Moscow region, 141700, Russia}}

\end{center}

\begin{center}\footnotesize{{\rm E-mails:}{\rm\
grekovandrew@mail.ru,\ shnbuz@gmail.com,\
zotov@mi-ras.ru}}\end{center}
%
%

 \begin{abstract}
We introduce a family of classical integrable systems describing
dynamics of $M$ interacting ${\rm gl}_N$ integrable tops. It extends
the previously known model of interacting elliptic tops. Our
construction is based on the ${\rm GL}_N$ $R$-matrix satisfying the
associative Yang-Baxter equation. The obtained systems can be
considered as extensions of the spin type Calogero-Moser models with
(the classical analogues of)  anisotropic spin exchange operators
given in terms of the $R$-matrix data. In $N=1$ case the spin
Calogero-Moser model is reproduced. Explicit expressions for ${\rm
gl}_{NM}$-valued Lax pair with spectral parameter and its classical
dynamical $r$-matrix are obtained. Possible applications
 are briefly
discussed.
 \end{abstract}

\newpage
\tableofcontents

\section{Introduction}
\setcounter{equation}{0}


In this paper we describe the classical integrable ${\rm gl}_{NM}$
model given by the Hamiltonian of the following form:
 \beq\label{t01}
 \begin{array}{c}
  \displaystyle{
 \mH=\sum\limits_{i=1}^M \frac{p_i^2}{2}+\sum\limits_{i=1}^M
 H^{\hbox{\tiny{top}}}(\mS^{ii})+\frac12\sum\limits_{i,j:\, i\neq j}^M
 \mU(\mS^{ij},\mS^{ji},q_i-q_j)\,,
 }
 \end{array}
 \eq
 where $p_i$ and $q_j$ are the canonical variables:
 \beq\label{t02}
 \begin{array}{c}
  \displaystyle{
 \{p_i,q_j\}=\delta_{ij}\,,\quad \{p_i,p_j\}=\{q_i,q_j\}=0\,,\quad
 i,j=1...M\,.
 }
 \end{array}
 \eq
 For all $i,j=1...M$ $\mS^{ij}$ are $N\times N$ matrices of
 ''classical spin''
variables, i.e.
 \beq\label{t03}
 \begin{array}{c}
  \displaystyle{
 \mS^{ij}=\sum\limits_{a,b=1}^N \mS^{ij}_{ab}\,e_{ab}\in \Mat\,,
 }
 \end{array}
 \eq
 where $\{e_{ab}\,,a,b=1...N\}$ is the standard basis in $\Mat$.
 They are naturally arranged into $NM\times NM$ block-matrix $\mS$:
 \beq\label{t04}
 \begin{array}{c}
  \displaystyle{
 \mS=\sum\limits_{i,j=1}^M E_{ij}\otimes \mS^{ij}
 =\sum\limits_{i,j=1}^M\sum\limits_{a,b=1}^N \mS^{ij}_{ab}\,E_{ij}\otimes e_{ab}\in \MatNM\,,
 }
 \end{array}
 \eq
  where $\{E_{ij}\,,i,j=1...M\}$ is the standard basis in $\MatM$.
The Poisson structure is given by the Poisson-Lie brackets on
 ${\rm gl}^*_{NM}$ Lie coalgebra:
  \beq\label{t05}
  \begin{array}{c}
  \displaystyle{
 \{\mS^{ij}_{ab},\mS^{kl}_{cd}\}=\mS^{kj}_{cb}\,\delta^{il}\,\delta_{ad}-\mS^{il}_{ad}\,\delta^{kj}\,\delta_{bc}\,.
 }
  \end{array}
 \eq

 \paragraph{Integrable tops.} In order to clarify the structure of the Hamiltonian (\ref{t01}) consider the case
 $M=1$. Then the last term in (\ref{t01}) is absent, and we are left
 with a free particle (with momenta $p_1$) and the Hamiltonian
 $H^{\hbox{\tiny{top}}}(\mS^{11})$ of integrable top of Euler-Arnold type
 \cite{Arnold}. Here we deal with the models admitting
 the Lax pairs with spectral parameter on elliptic curves
 \cite{Skl1}.
 The general form for equations of motion (for the top like models) is
 \beq\label{t001}
 \begin{array}{c}
  \displaystyle{
\dot {S}=[{S}, {J}({S})]\,,
 }
 \end{array}
 \eq
 where $S\in\Mat$ is the matrix of dynamical variables, while the
 inverse inertia tensor $J$ is a linear map
  \beq\label{t002}
  J(S)=\sum\limits_{i,j,k,l=1}^N J_{ijkl}\,e_{ij}\,S_{lk}\in\Mat
  \eq
 In the general case the model (\ref{t001}) is not integrable. It is
 integrable for some special $J(S)$ only.
More precisely, here we consider special tops, which were described
in
 \cite{LOZ,ZS}, \cite{KrZ}, \cite{AASZ,LOZ8} for elliptic, trigonometric and rational cases respectively.
 All of them can be written \cite{LOZ8,LOZ16} in the $R$-matrix form based on a quantum
 ${\rm GL}_N$ $R$-matrix (in the fundamental representation) satisfying
 the associative Yang-Baxter equation \cite{FK,Pol}:
  \beq\label{t003}
  R^\hbar_{12}(q_{12})
 R^{\eta}_{23}(q_{23})=R^{\eta}_{13}(q_{13})R_{12}^{\hbar-\eta}(q_{12})+
 R^{\eta-\hbar}_{23}(q_{23})R^\hbar_{13}(q_{13})\,,\quad
 q_{ab}=q_a-q_b\,.
  \eq
 Having solution of (\ref{t003}) with some additional properties (see the next Section)
  the inverse inertia tensor comes
 from the term $m_{12}(z)$ in the classical limit expansion:
  \beq\label{t004}
  \begin{array}{c}
      \displaystyle{
R^\hbar_{12}(z)=\frac{1}{\hbar}\,1_N\otimes 1_N+r_{12}(z)+\hbar\,
m_{12}(z)+O(\hbar^2) }\,.
  \end{array}
  \eq
 Namely, for
  \beq\label{t005}
  \begin{array}{c}
      \displaystyle{
 m_{12}(z)=\sum\limits_{i,j,k,l=1}^N m_{ijkl}(z)\, e_{ij}\otimes e_{kl}
  }
  \end{array}
  \eq
the components of $J$ are
  \beq\label{t006}
  \begin{array}{c}
      \displaystyle{
 J_{ijkl}=m_{ijkl}(0)\,,
  }
  \end{array}
  \eq
 that is
  \beq\label{t007}
  \begin{array}{c}
      \displaystyle{
 J(S)=\tr_2(m_{12}(0)S_2)\,,\qquad S_2=1_N\otimes S\,.
  }
  \end{array}
  \eq
 The Hamiltonian of the model is of the form:
  \beq\label{t008}
  \begin{array}{c}
     \displaystyle{
 H^{\hbox{\tiny{top}}}(S)=\frac{1}{2}\,\tr(SJ(S))=\frac{1}{2}\,\tr_{12}(m_{12}(0)S_1S_2)\,,\qquad
 S_1=S\otimes 1_N\,.
 }
  \end{array}
  \eq
  This expression enters (\ref{t01}). The phase space of the model
  is a coadjoint orbit
  \beq\label{t009}
  \begin{array}{c}
     \displaystyle{
 \mM^{\hbox{\tiny{top}}}=\mO_N
 }
  \end{array}
  \eq
  of ${\rm GL}_N$ Lie group, i.e. the space
  spanned by $S_{ij}$ with some fixed eigenvalues of matrix  $S$ (or the Casimir functions $C_k=\tr
  S^k$). Its dimension depends on the eigenvalues. The
  minimal orbit $\mO_N^{\hbox{\tiny{min}}}$ corresponds to $N-1$ coincident
  eigenvalues, i.e  the matrix $S$ (up to a matrix proportional to identity matrix) is of rank one:
  \beq\label{t0091}
  \begin{array}{c}
     \displaystyle{
 \dim\mO_N^{\hbox{\tiny{min}}}=2(N-1)\,.
 }
  \end{array}
  \eq
 The Lax pair is given in the Appendix C.

 \paragraph{Spin Calogero-Moser model.} In the case $N=1$ the second term in (\ref{t01}) is trivial, and the last
 one boils down to the spin Calogero-Moser model \cite{GH,BAB}:
  \beq\label{t06}
  \begin{array}{c}
  \displaystyle{
H^{\hbox{\tiny{spin}}}=\sum\limits_{i=1}^M\frac{p_i^2}{2}-\sum\limits_{i>j}^M
S_{ij}S_{ji}E_2(q_i-q_j)\,,
 }
 \end{array}
 \eq
where $E_2(q)$ is the second Eisenstein function (\ref{a04}). Some
details of the spin Calogero-Moser model are given in the Appendix
B. Let us only remark here that the model (\ref{t06}) is integrable
through the Lax representation and the classical $r$-matrix
structure on the constraints
  \beq\label{t07}
  \begin{array}{c}
  \displaystyle{
 S_{ii}=\nu\ \hbox{for all}\ i=1...M
 }
 \end{array}
 \eq
 supplemented by some gauge fixation conditions generated by the
 coadjoint action of the Cartan subgroup
  ${\mathfrak H}_M\subset{\rm GL}_M$. That is the phase space of the model is given by
  \beq\label{t08}
  \begin{array}{c}
  \displaystyle{
 {\mathcal M}^{\hbox{\tiny{spin}}}=T^*{\mathfrak h}_M\times
  {\mathcal O_M}//{\mathfrak H}_M\,,
 }
 \end{array}
 \eq
where ${\mathfrak h}_M=Lie({\mathfrak H}_M)$ is the Lie algebra of
${\mathfrak H}_M$, and ${\mathcal O}_M$ is an orbit of the coadjoint
action of ${\rm GL}_M$. The first factor in (\ref{t08}) describes
the many-body degrees of freedom (\ref{t02}), and the second factor
describes the "classical spin" variables. In the general case the
spin variables can be parameterized by the set of canonically
conjugated variables:
  \beq\label{t081}
  \begin{array}{c}
  \displaystyle{
S_{ij}=\sum\limits_{a=1}^N\xi^i_a\eta^j_a\,,
  }
  \\ \ \\
  \displaystyle{
 \{\xi^i_a,\eta^j_b\}=\delta_{ab}\delta_{ij}\,,\quad
 i,j=1...M\,,\quad a,b=1...N\,.
 }
 \end{array}
 \eq
 The Poisson structure (\ref{a25}) is reproduced in this way. Using
 these notations it is easy to see that
  \beq\label{t082}
  \begin{array}{c}
  \displaystyle{
 S_{ij}S_{ji}=\sum\limits_{a,b=1}^N\xi^i_a\eta^j_a\xi^j_b\eta^i_b=\tr(\mS^{ii}\mS^{jj})\,,
 }
 \end{array}
 \eq
 and the potential in the Hamiltonian (\ref{t06}) takes the form
  \beq\label{t083}
  \begin{array}{c}
  \displaystyle{
 \mV^{\hbox{\tiny{spin}}}(\mS^{ii},\mS^{jj},q_{ij})=-\tr(\mS^{ii}\mS^{jj})E_2(q_i-q_j)\,.
 }
 \end{array}
 \eq
 Below we construct anisotropic (in $\Mat$ space) generalizations of (\ref{t083}).

In the special case, when the matrix of spin variables $S$ is of
rank 1 (it is the minimal $\mO_M^{\hbox{\tiny{min}}}$ orbit
(\ref{t0091}))
  \beq\label{t09}
  \begin{array}{c}
  \displaystyle{
 S_{ij}=\xi_i\eta_j
 }
 \end{array}
 \eq
the reduction with respect to the action of ${\mathfrak H}_M$ leads
to the spinless Calogero-Moser (CM) model \cite{Ca,Krich1} since the
second factor in (\ref{t08}) become trivial. Indeed, plugging
(\ref{t09}) into (\ref{t06}) and using (\ref{t07}) we get
  \beq\label{t10}
  \begin{array}{c}
  \displaystyle{
H^{\hbox{\tiny{spin}}}=\sum\limits_{i=1}^M\frac{p_i^2}{2}-\nu^2\sum\limits_{i>j}^M
E_2(q_i-q_j)\,.
 }
 \end{array}
 \eq
 The spinless Calogero-Moser models are
gauge equivalent to the special top with the minimal orbit
(\ref{t0091}). See \cite{LOZ,KrZ,AASZ} for details.

 \paragraph{Interacting tops.} Turning back to the ${\rm gl}_{NM}$ model
 (\ref{t01}) consider the special case when the matrix $\mS$ is of
 rank 1:
  \beq\label{t201}
  \begin{array}{c}
  \displaystyle{
 \mS^{ij}_{ab}=\xi^i_a\eta^j_b\,.
 }
 \end{array}
 \eq
  We will see that in this case the last term in (\ref{t01}) is
  rewritten in the form
 \beq\label{t202}
 \begin{array}{c}
  \displaystyle{
 \mU(\mS^{ij},\mS^{ji},q_i-q_j)=\mV(\mS^{ii},\mS^{jj},q_i-q_j)\,,
 }
 \end{array}
 \eq
  and the Hamiltonian (\ref{t01}) acquires the form
  \beq\label{t203}
  \begin{array}{c}
  \displaystyle{
 \mH^{\hbox{\tiny{tops}}}=\sum\limits_{i=1}^M \frac{p_i^2}{2}+\sum\limits_{i=1}^M
 H^{\hbox{\tiny{top}}}(\mS^{ii})+\frac12\sum\limits_{i,j:\, i\neq j}^M
 \mV(\mS^{ii},\mS^{jj},q_i-q_j)\,.
 }
 \end{array}
 \eq
 It describes mechanics of $M$ interacting integrable ${\rm gl}_N$
 tops. The Hamiltonian of (\ref{t203}) type was introduced by A.P. Polychronakos
 \cite{Polych} from his study of matrix models. Then the elliptic version of model (\ref{t01}) and (\ref{t203}) was described as
 ${\rm gl}_{NM}$
 Hitchin system \cite{ZL,LOSZ4,ZS} (see some details in Section \ref{ex}), and (\ref{t01}) was also generalized for arbitrary
 complex Lie group \cite{LOSZ}.


Similarly to the spin Calogero-Moser model  the general model
(\ref{t01}) requires additional constraints (cf. (\ref{t07}))
  \beq\label{t204}
  \begin{array}{c}
  \displaystyle{
 \tr(\mS^{ii})=\nu\ \hbox{for all}\ i=1...M\,.
 }
 \end{array}
 \eq
 They should be supplied with some gauge fixation conditions
 generated by the coadjoint action of ${\mathfrak H}_{NM}'\subset{\mathfrak H}_{NM}$ --
 subgroup of the Cartan subgroup ${\mathfrak H}_{NM}\subset{\rm
 GL}_{NM}$ with elements of the form $\sum\limits_{i=1}^M h_i\, E_{ii}\otimes
 1_N$. Together with (\ref{t204}) the gauge fixation conditions are
 the second class constraints, and one can perform the Dirac
 reduction procedure to compute the final Poisson structure
 starting from the linear one (\ref{t05}). The phase space of the
 general model (\ref{t01}) is of the from:
  \beq\label{t205}
  \begin{array}{c}
  \displaystyle{
 {\mathcal M}^{\hbox{\tiny{gen}}}=T^*{\mathfrak h}_{NM}'\times
  {\mathcal O_{NM}}//{\mathfrak H}_{NM}'\,,\qquad
   {\mathfrak h}_{NM}'=Lie({\mathfrak H}_{NM}')\,.
 }
 \end{array}
 \eq
 For the interacting tops case (\ref{t201})-(\ref{t203})
the orbit
 $\mO_{NM}$ becomes $\mO_{NM}^{\hbox{\tiny{min}}}$. Then the phase
 space
  \beq\label{t206}
  \begin{array}{c}
  \displaystyle{
 {\mathcal M}^{\hbox{\tiny{tops}}}=T^*{\mathfrak h}_{NM}'\times
  \mathcal O_{NM}^{\hbox{\tiny{min}}}//{\mathfrak H}_{NM}'
 }
 \end{array}
 \eq
 has dimension $2NM$, while its ''spin part'' is of dimension
  \beq\label{t207}
  \begin{array}{c}
  \displaystyle{
 \dim
  \Big(\mathcal O_{NM}^{\hbox{\tiny{min}}}//{\mathfrak
  H}_{NM}'\Big)=2NM-2M\,.
 }
 \end{array}
 \eq
 A brief summary of the described models is given in the following
 scheme:
  \beq
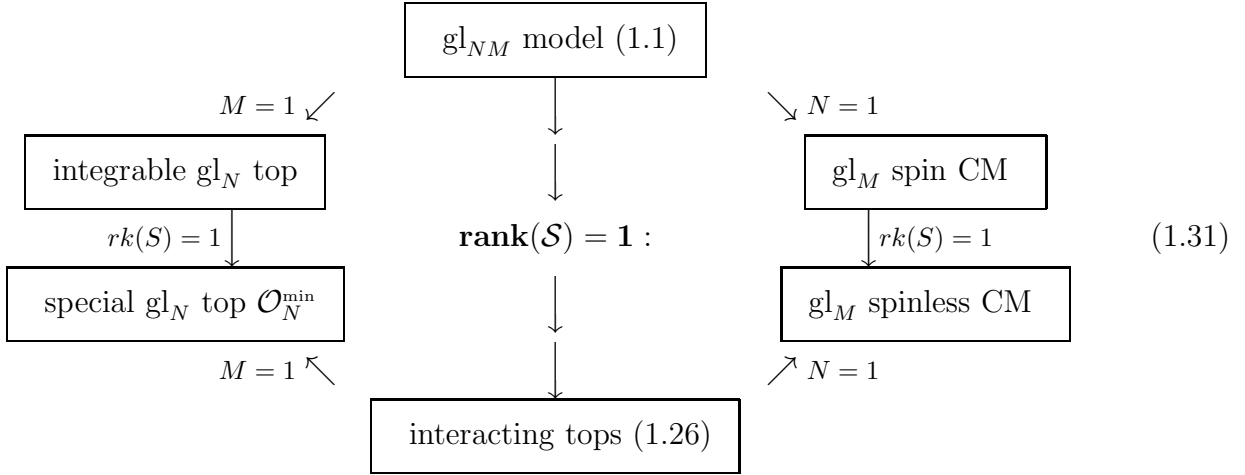
\label{q303}
   \begin{array}{ccc}
   & \fbox{$\phantom{\Big(}$ ${\rm gl}_{NM}$ model (\ref{t01})$\phantom{\Big(}$} &
   \\
  $\qquad\qquad\qquad\phantom{\Big(}$ \hbox{\footnotesize{$M=1$}}\swarrow &  \Big\downarrow & \searrow
  \hbox{\footnotesize{$N=1$}} $\phantom{\Big(}\qquad\qquad\qquad$
      \\
    \fbox{$\phantom{\Big(}$integrable ${\rm gl}_N$ top$\phantom{\Big(}$} &  \Big\downarrow &
    \fbox{$\phantom{\Big(}$${\rm gl}_M$ spin CM $\phantom{\Big(}$}
         \\
  \hbox{\footnotesize{$rk(S)=1$}} \Big\downarrow &  $\phantom{\Big(}$\bf{rank(\mS)=1:} $\phantom{\Big(}$ &
   \Big\downarrow\hbox{\footnotesize{$rk(S)=1$}}
            \\
    \fbox{$\phantom{\Big(}$ special ${\rm gl}_N$ top $\mO_N^{\hbox{\tiny{min}}}$$\phantom{\Big(}$} &  \Big\downarrow &
    \fbox{$\phantom{\Big(}$${\rm gl}_M$ spinless CM $\phantom{\Big(}$}
    \\
  $\qquad\qquad\qquad\phantom{\Big(}$ \hbox{\footnotesize{$M=1$}}\nwarrow &  \Big\downarrow & \nearrow
  \hbox{\footnotesize{$N=1$}} $\phantom{\Big(}\qquad\qquad\qquad$
  \\
 & \fbox{$\phantom{\Big(}$  interacting tops (\ref{t203})$\phantom{\Big(}$} &
   \end{array}
   \eq


\paragraph{Purpose of the paper} is to describe a family of the models (\ref{t01}) and (\ref{t203})
in terms of $R$-matrices satisfying the associative Yang-Baxter
equation (\ref{t003}). We give explicit formulae for $NM\times NM$
Lax pair with spectral parameter (see the next Section) and compute
the Hamiltonians (\ref{t01}) and (\ref{t203}). As a result we obtain
the potentials
  \beq\label{t208}
  \begin{array}{c}
      \displaystyle{
 \mU(\mS^{ij},\mS^{ji},q_i-q_j)=\tr_{12}\Big(
 \p_{q_i}r_{21}(q_{ij}) P_{12}\,\mS_1^{ij}\mS_2^{ji}\Big)
  }
  \end{array}
  \eq
for the general model (\ref{t01}) and
  \beq\label{t209}
  \begin{array}{c}
      \displaystyle{
 \mV(\mS^{ii},\mS^{jj},q_i-q_j) =\tr_{12}\Big(\p_{q_i}r_{12}(q_{ij})\mS^{ii}_1\mS^{jj}_2\Big)
 }
  \end{array}
  \eq
for the model of interacting tops (\ref{t203}). Notice that in the
simplest case related to the rational Yang's XXX $R$-matrix
  \beq\label{t20901}
  \begin{array}{c}
      \displaystyle{
  R_{12}^z(q_{ij})=\frac{1_N\otimes 1_N}{z}+\frac{P_{12}}{q_i-q_j}
 }
  \end{array}
  \eq
 we get just the spin Calogero-Moser model written in terms of
 matrix variables:
  \beq\label{t20902}
  \begin{array}{c}
      \displaystyle{
 \mV=-\frac{ \tr(\mS^{ii}\mS^{jj}) }{(q_i-q_j)^2}\,.
  }
  \end{array}
  \eq

Next, we proceed to the classical (dynamical) $r$-matrix. It is
similar to the one for the spin Calogero-Moser case \cite{BAB} but
this time its matrix elements are $R$-matrices themselves. The
classical exchange relations are verified directly. This guarantees
the Poisson commutativity of the Hamiltonians generated by the Lax
matrix.

The answers (\ref{t208}) and (\ref{t209}) depend on the classical
$r$-matrix, which appears from the quantum one in the limit
(\ref{t004}). The quantum $R$-matrix enters the higher Hamiltonians.
It should satisfy a set of properties which we discuss in the next
Section. The most general $R$-matrix satisfying all the required
properties is the elliptic Baxter-Balavin's one. In this case the
the integrable models are known. They were first described by
Polychronakos in \cite{Polych} and later reproduced as Hitchin type
systems on the bundles with nontrivial characteristic classes in
\cite{ZL,LOSZ4}.

The family of the obtained models includes new integrable systems in
the trigonometric and rational cases. While the quantization of the
potential $\mV$ from (\ref{t20902}) is given by isotropic spin
exchange operator $\hat\mV=-P_{ij}/(q_i-q_j)^2$, the obtained
general answer (\ref{t208})-(\ref{t209}) leads to the anisotropic
potentials. An example of such anisotropic extension to the spin
(trigonometric) Calogero-Moser-Sutherland model was first suggested
by Hikami and Wadati \cite{Hikami} at quantum level. From the point
of view
 of (\ref{t209}) their answer corresponds to the ${\rm gl}_2$ XXZ
$r$-matrix. At the same time the set of trigonometric $R$-matrices
satisfying the required properties is much lager \cite{AHZ,Sch}, and
all these  $R$-matrices can be used for construction of the
integrable tops \cite{KrZ}. The results of the present paper are
also valid for all these cases. An example based on the ${\rm gl}_2$
7-th vertex deformation of the XXZ $R$-matrix is given Section 4.
Similarly, in the rational case the admissible $R$-matrices includes
not only the Yang's $R$-matrix (\ref{t20901}) but also its
deformations such as 11-vertex $R$-matrix \cite{Cherednik2} and its
higher rank versions \cite{LOZ8}. An example related to 11-vertex
$R$-matrix is given in Section 4.


Possible applications of the described models are discussed in the
end. Namely, we argue that the obtained models can be used for
construction of higher Hamiltonians for the anisotropic
generalizations of the Haldane-Shastry-Inozemtsev long-range spin
chains. The latter is important for the proof of integrability of
these chains, which still remains an open problem.


\section{Lax equations}
\setcounter{equation}{0} In this Section we construct the $NM\times
NM$ Lax pair $\mL(z),\mM(z)$ satisfying the Lax equations

  \beq\label{t3000}
  \begin{array}{c}
      \displaystyle{
 {\dot \mL}(z)=[\mL(z),\mM(z)]
}
  \end{array}
  \eq
for the model (\ref{t01}). Our construction is based on ${\rm GL}_N$
$R$-matrix -- solution of the associative Yang-Baxter equation
(\ref{t003}). Besides (\ref{t003}) the $R$-matrix should also
satisfy a set of properties.

\subsection{$R$-matrix properties.}\label{S21} We consider $R$-matrices satisfying (\ref{t003}) and
(\ref{t004}). Let us also impose the following set of conditions for
${\rm GL}_N$ $R$-matrices under consideration:

\noindent \underline{Expansion near $z=0$:}
  \beq\label{t300}
  \begin{array}{c}
      \displaystyle{
R^\h_{12}(z)=\frac{1}{z}\,P_{12}+R^{\h,(0)}_{12}+zR^{\h,(1)}_{12}+O(z^2)\,,
}
  \end{array}
  \eq
 Also,
  \beq\label{q402}
  \begin{array}{c}
      \displaystyle{
 R^{z,(0)}_{12}=\frac{1}{z}\,1_N\otimes
 1_N+r^{(0)}_{12}+O(z)\,,\qquad
 r_{12}(z)=\frac{1}{z}\,P_{12}+r^{(0)}_{12}+z\,r^{(1)}_{12}+O(z^2)\,.
 }
  \end{array}
  \eq
\noindent \underline{Skew-symmetry:}\footnote{$P_{12}$ entering
(\ref{t301}) is the permutation operator, $(P_{12})^2=1_N\otimes
1_N$.}
  \beq\label{t301}
  \begin{array}{c}
  \displaystyle{
 R^\hbar_{12}(z)=-R_{21}^{-\hbar}(-z)=-P_{12}R_{12}^{-\hbar}(-z)P_{12}\,,
 \qquad
 P_{12}=\sum\limits_{i,j=1}^N E_{ij}\otimes E_{ji}\,.
 }
 \end{array}
 \eq
\noindent  \underline{Unitarity:}
 \beq\label{t302}
   \begin{array}{c}
 \displaystyle{
R^\hbar_{12}(z) R^\hbar_{21}(-z) = f^\hbar(z)\,\,1_N\otimes
1_N\,,\qquad f^\hbar(z)=\wp(\hbar)-\wp(z)\,.
 }
  \end{array}
  \eq
 We are also going to use \underline{the Fourier
 symmetry}:
%
 \beq\label{t303}
   \begin{array}{c}
 \displaystyle{
R^\hbar_{12}(z) P_{12}=R^z_{12}(\hbar)
 }
  \end{array}
  \eq
  It is not necessary but convenient property.
 The following relations on the coefficients of expansions (\ref{t004}) and (\ref{t300}) follow from the
skew-symmetry:
  \beq\label{t304}
  \begin{array}{l}
  r_{12}(z)=-r_{21}(-z)\,,\qquad
  m_{12}(z)=m_{21}(-z)\,,\\ \ \\
      R^{\h,(0)}_{12}=-R^{-\h,(0)}_{21}\,,\qquad
  r_{12}^{(0)}=-r_{21}^{(0)}\,.
  \end{array}
  \eq
Similarly, from the Fourier symmetry we have (see details in
\cite{Z2}):
  \beq\label{t305}
  \begin{array}{l}
  R^{z,(0)}_{12}=r_{12}(z)P_{12}\,,\\ \ \\
      R^{z,(1)}_{12}=m_{12}(z)P_{12}\,,\\ \ \\
  r_{12}^{(0)}=r_{12}^{(0)}P_{12}\,.
  \end{array}
  \eq
In what follows we use special notation for the $R$-matrix
derivative:
  \beq\label{t521}
  \begin{array}{c}
  \displaystyle{
 F^z_{12}(q)=\p_q R_{12}^z(q)
 }
  \end{array}
 \eq
It is the $R$-matrix analogue of the function (\ref{a05}) entering
the $M$-matrix of the spin Calogero-Moser model (\ref{a36}) likewise
$R$-matrix itself is a matrix analogue of the Kronecker function
(\ref{a01}) due to similarity of (\ref{a06}) and (\ref{t003}).
 See \cite{LOZ9}.
 Then from the classical limit (\ref{t004}) we have
  \beq\label{t522}
  \begin{array}{c}
  \displaystyle{
 F^0_{12}(q)=\p_q R_{12}^z(q)\left.\right|_{z=0}=\p_q r_{12}(q)\,.
 }
  \end{array}
 \eq
 The latter is the $R$-matrix analogue of the function $-E_2(q)$ (\ref{a13})
 entering the Calogero-Moser potential.
Notice also that $F^0_{12}(q)=F^0_{21}(-q)$ due to (\ref{t304}).
From (\ref{t521}) and (\ref{t300}) the local expansion near $q=0$ is
as follows
  \beq\label{t5221}
  \begin{array}{c}
  \displaystyle{
 F^z_{12}(q)=-\frac{1}{q^2}\,P_{12}+R^{z,(1)}_{12}+O(q)
 }
  \end{array}
 \eq
and, therefore,
  \beq\label{t5222}
  \begin{array}{c}
  \displaystyle{
 F^0_{12}(q)=-\frac{1}{q^2}\,P_{12}+R^{z,(1)}_{12}\left.\right|_{z=0}+O(q)\stackrel{(\ref{t305})}{=}
 -\frac{1}{q^2}\,P_{12}+m_{12}(0)P_{12}+O(q)\,.
 }
  \end{array}
 \eq
On the other hand
  \beq\label{t5223}
  \begin{array}{c}
  \displaystyle{
 F^0_{12}(q)\stackrel{(\ref{t522})}{=}\p_q
 r_{12}(q)\stackrel{(\ref{q402})}{=}-\frac{1}{q^2}\,P_{12}+r_{12}^{(1)}+O(q)\,.
 }
  \end{array}
 \eq
 From (\ref{t5222}) and  (\ref{t5223}) we conclude that
  \beq\label{t5224}
  \begin{array}{c}
  \displaystyle{
 r_{12}^{(1)}=m_{12}(0)P_{12}\,.
 }
  \end{array}
 \eq
 In the elliptic case the set of properties is fulfilled by the
 Baxter-Belavin \cite{Belavin} $R$-matrix (\ref{c61}). A family of
 trigonometric $R$-matrices include the XXZ 6-vertex one, its
 7-vertex deformation \cite{Cherednik2} and ${\rm GL}_N$
 generalizations \cite{AHZ,Sch}. See a brief review and applications
 to integrable tops in \cite{KrZ}. The rational $R$-matrices
 possessing the properties are the XXX Yang's $R$-matrix, its
 11-vertex deformation \cite{Cherednik2} and higher rank analogues
 obtained from the elliptic case by special limiting procedure
 \cite{Smirnov}. The final answer for such $R$-matrix was obtained in \cite{LOZ8}
 through the gauge equivalence between the relativistic top with
 minimal orbit and the rational Ruijsenaars-Schneider model.


 \subsection{Lax pair and equations of motion}
Using coefficients of the expansion of the ${\rm GL}_N$ $R$-matrix
near $z=0$
we define $NM\times NM$ Lax pair
  \beq\label{t50}
  \begin{array}{c}
  \displaystyle{
 L(z)=\sum\limits_{i,j=1}^M E_{ij}\otimes L^{ij}(z)\,,\quad\quad
 L^{ij}(z)\in{\rm Mat}_{N}\quad\quad L(z)\in{\rm Mat}_{NM}\,,
 }
  \end{array}
 \eq
  \beq\label{t51}
  \begin{array}{c}
  \displaystyle{
 L^{ij}(z)=
 \delta_{ij}\Big(p_i1_N+\tr_2(\mS^{\,ii}_2\,R_{12}^{z,(0)}P_{12})\Big)
  +(1-\delta_{ij})\,\tr_2(\mS^{\,ij}_2\,R^z_{12}(q_{ij})P_{12})\,.
 }
  \end{array}
 \eq
and similarly for $M^{ij}(z)\in{\rm Mat}_{N}$
  \beq\label{t52}
  \begin{array}{c}
  \displaystyle{
 M^{ij}(z)=
 \delta_{ij}\,\tr_2(\mS^{\,ii}_2\,R_{12}^{z,(1)}P_{12})
  +(1-\delta_{ij})\,\tr_2(\mS^{\,ij}_2\,F^z_{12}(q_{ij})P_{12})\,.
 }
  \end{array}
 \eq
 where the entries are defined from (\ref{t300}) and (\ref{t521}).
 The tensor notations are similar to those used in
 (\ref{c409})-(\ref{c412}).

 \begin{predl}
 Consider an $R$-matrix satisfying the associative Yang-Baxter
 equation (\ref{t003}), the classical limit (\ref{t004}) and the set
 of properties from the previous paragraph.
 Then the Lax equation (\ref{t3000}) holds true for the Lax pair (\ref{t50})-(\ref{t52})
  on the constraints
  \beq\label{t523}
  \begin{array}{c}
  \displaystyle{
 \tr\Big(\mS^{ii}\Big)=\hbox{const}\,,\ \forall i
 }
  \end{array}
 \eq
 (cf. (\ref{t23})) and provides the following equations of motion
for off-diagonal $N\times N$ blocks of $\mS$:
  \beq\label{t524}
  \begin{array}{c}
  \displaystyle{
 {\dot
 \mS}^{ij}=\sum\limits_{k:k\neq i,j}^M
 \Big(\mS^{ik}\tr_{2}(\mS^{kj}_2F_{12}^0(q_{kj})P_{12})-\tr_2(\mS_2^{ik}F_{12}^0(q_{ik})P_{12})\mS^{kj}\Big)+
 }
 \\ \ \\
 \displaystyle{
  +\mS^{ii}\tr_2(\mS_2^{ij}F^0_{12}(q_{ij})P_{12})-\tr_2(\mS^{ii}_2m_{12}(0))\mS^{ij}-
 }
  \\ \ \\
 \displaystyle{
  -\tr_2(\mS_2^{ij}{ F}^0_{12}(q_{ij})P_{12})\mS^{jj}+\mS^{ij}\tr_2(\mS_2^{jj}m_{12}(0))\,,
 }
  \end{array}
 \eq
  for diagonal $N\times N$ blocks of $\mS$:
  \beq\label{t525}
  \begin{array}{c}
  \displaystyle{
 {\dot \mS}^{ii}=[\mS^{ii},\tr_2(m_{12}(0)\mS^{ii}_2)]+\sum\limits_{k:k\neq
 i}^M\Big( \mS^{ik}\tr_2(\mS_2^{ki}F_{21}^0(q_{ik})P_{12})-\tr_2(\mS^{ik}_2F_{12}^0(q_{ik})P_{12})\mS^{ki}
 \Big)\,,
 }
  \end{array}
 \eq
  and for momenta:
  \beq\label{t526}
  \begin{array}{c}
  \displaystyle{
 {\dot p}_i=-\sum\limits_{k:k\neq
 i}^M\tr_{23}\Big(\p_{q_i}F^0_{32}(q_{ik})P_{23}\,\mS^{ik}_2\mS^{ki}_3\Big)\,.
 }
  \end{array}
 \eq
 \end{predl}
\noindent\underline{\em{Proof:}}\quad
   We imply $p_i={\dot q}_i$ in the formulae above. This follows from
 the Hamiltonian description, which is given in the next paragraph.

\noindent {\bf 1.} Let us begin with the non-diagonal blocks.
Consider the one numbered
 $ij$ ($i\neq j$). The l.h.s. of the Lax equations reads
  \beq\label{t527}
  \begin{array}{c}
  \displaystyle{
 \hbox{l.h.s.}={\dot \mL}^{ij}(z)=\tr_2({\dot
 \mS}_2^{ij}R_{12}^z(q_{ij})P_{12})+\tr_2({\mS}_2^{ij}F_{12}^z(q_{ij})P_{12})({\dot q}_i-{\dot
 q}_j)\,.
 }
  \end{array}
 \eq
 The r.h.s. of the Lax equation is as follows:
  \beq\label{t528}
  \begin{array}{c}
  \displaystyle{
 \hbox{r.h.s.}=\mL^{ij}\mM^{jj}-\mM^{ii}\mL^{ij}+\mL^{ii}\mM^{ij}-\mM^{ij}\mL^{jj}
 +\sum\limits_{k:k\neq i,j}^M \Big(\mL^{ik}\mM^{kj}-\mM^{ik}\mL^{kj}\Big)\,.
 }
  \end{array}
 \eq
 The last sum is computed using identity
  \beq\label{t529}
  \begin{array}{c}
  \displaystyle{
  R_{12}^z(x)F_{23}^z(y)-F_{12}^z(x)R_{23}^z(y)=F^0_{23}(y)R^z_{13}(x+y)-R^z_{13}(x+y)F_{12}^0(x)\,,
 }
  \end{array}
 \eq
 which follows from (\ref{t003}). It is the $R$-matrix analogue of
 (\ref{a07}). In its turn (\ref{a07}) is the key tool underlying ansatz for the Lax pairs with
 spectral parameter \cite{Krich1}. For $k\neq i,j$ we have
  \beq\label{t530}
  \begin{array}{c}
  \displaystyle{
 \mL^{ik}\mM^{kj}-\mM^{ik}\mL^{kj}=
 }
 \\ \ \\
   \displaystyle{
 =\tr_{23}(R_{12}^z(q_{ik})P_{12}\mS_2^{ik}F_{13}^z(q_{kj})P_{13}\mS_3^{kj})
  -\tr_{23}(F_{12}^z(q_{ik})P_{12}\mS_2^{ik}R_{13}^z(q_{kj})P_{13}\mS_3^{kj})=
 }
  \\ \ \\
   \displaystyle{
 =\tr_{23}\Big(\Big(R_{12}^z(q_{ik})F_{23}^z(q_{kj})
 -F_{12}^z(q_{ik})R_{23}^z(q_{kj})\Big)P_{12}P_{13}\,\mS_2^{ik}\mS_3^{kj}\Big)\stackrel{(\ref{t529})}{=}
 }
   \\ \ \\
   \displaystyle{
 =\tr_{23}\Big(\Big(F^0_{23}(q_{kj})R^z_{13}(q_{ij})-R^z_{13}(q_{ij})F_{12}^0(q_{ik})\Big)P_{12}P_{13}\,\mS_2^{ik}\mS_3^{kj}\Big)=
 }
    \\ \ \\
   \displaystyle{
 =\tr_{23}\Big(R_{12}^z(q_{ij})P_{12}\Big(\mS_2^{ik}\mS_3^{kj}F^0_{23}(q_{kj})P_{23}-F^0_{23}(q_{ik})P_{23}\,\mS_3^{ik}\mS_2^{kj}\Big)\Big)\,.
 }
  \end{array}
 \eq
 This expression provides the upper line in the equations of motion
(\ref{t524}). To proceed we need degenerations of the identity
(\ref{t529}) when $y\rightarrow 0$. It comes from the expansions
(\ref{t300}), (\ref{t5221}) and (\ref{t5223}):
  \beq\label{t531}
  \begin{array}{c}
  \displaystyle{
  R_{12}^z(x)R_{23}^{z,(1)}-F_{12}^z(x)R_{23}^{z,(0)}=r_{23}^{(1)}R_{13}^z(x)-R_{13}^z(x)F_{12}^0(x)-\frac{1}{2}\,P_{23}\,\p^2_x
  R^z_{13}(x)\,.
 }
  \end{array}
 \eq
In the same way in the limit $x\rightarrow 0$ (\ref{t529}) takes the
form
  \beq\label{t532}
  \begin{array}{c}
  \displaystyle{
  R_{12}^{z,(0)}F_{23}^z(y)-R_{12}^{z,(1)}R_{23}^z(y)=F_{23}^0(y)R_{13}^z(y)-R_{13}^z(y)r_{12}^{(1)}+\frac{1}{2}\,\p^2_y
  R^z_{13}(y)P_{12}\,.
 }
  \end{array}
 \eq
 Similarly to the ordinary (spin) Calogero-Moser case the terms linear in
 momenta in the r.h.s. (\ref{t528}) $(p_i-p_j)\mM^{ij}$ are cancelled
 out by the last term in the l.h.s. of (\ref{t527}). Consider the first
 and the fourth terms from (\ref{t528}) without momenta. Using
 evaluations similar to (\ref{t530}) we get
  \beq\label{t533}
  \begin{array}{c}
  \displaystyle{
  \mL^{ij}\mM^{jj}-\mM^{ij}(\mL^{jj}-p_j1_N)=
 }
 \\ \ \\
   \displaystyle{
  =\tr_{23}\Big(\Big(R_{12}^z(q_{ij})R_{23}^{z,(1)}-F_{12}^z(q_{ij})R_{23}^{z,(0)}\Big)P_{12}P_{13}\,\mS_2^{ij}\mS_3^{jj}\Big)
  \stackrel{(\ref{t531})}{=}
 }
 \\ \ \\
   \displaystyle{
  =\tr_{23}\Big(\Big(r_{23}^{(1)}R_{13}^z(q_{ij})-R_{13}^z(q_{ij})F_{12}^0(q_{ij})-\frac{1}{2}\,P_{23}\,\p^2_{q_{i}}
  R^z_{13}(q_{ij})\Big)P_{12}P_{13}\,\mS_2^{ij}\mS_3^{jj}\Big)=
 }
  \\ \ \\
   \displaystyle{
  =\tr_{23}(R_{12}^z(q_{ij})P_{12}\,\mS_2^{ij}\mS^{jj}_3m_{23}(0))
  -\tr_{23}(R_{12}^z(q_{ij})P_{12}F^0_{23}(q_{ij})P_{23}\,\mS_3^{ij}\mS_2^{jj})-
 }
   \\ \ \\
   \displaystyle{
  -\frac{1}{2}\,\tr_{23}(\p^2_{q_{i}}
  R^z_{12}(q_{ij})P_{12}\mS_2^{ij}\mS^{jj}_3)\,,
 }
  \end{array}
 \eq
 where the relation (\ref{t5224}) was also used (for the first term in the answer). The first and the
 second terms in the obtained answer provide the last line in the
 equations of motion (\ref{t524}), while the last term in
 (\ref{t524}) is the ''unwanted term''.

 In the same way, using (\ref{t532}) one gets
  \beq\label{t534}
  \begin{array}{c}
  \displaystyle{
 (\mL^{ii}-p_i1_N)\mM^{ij}-\mM^{ii}\mL^{ij}=
 }
 \\ \ \\
   \displaystyle{
 =\tr_{23}(R_{12}^z(q_{ij})P_{12}\,\mS_2^{ii}\mS^{ij}_3F^0_{23}(q_{ij})P_{23})
 -\tr_{23}(R_{12}^z(q_{ij})P_{12}m_{23}(0)\mS_3^{ii}\mS^{ij}_2)+
 }
  \\ \ \\
   \displaystyle{
 +\frac{1}{2}\,\tr_{23}(\p^2_{q_{i}}
  R^z_{12}(q_{ij})P_{12}\mS_3^{ii}\mS^{ij}_2)\,.
 }
  \end{array}
 \eq
 Again, the first two terms provide an input to equations of motion
 -- the second line in (\ref{t524}). The last term is the ''unwanted
 term''. It is cancelled by the one from (\ref{t533}) after taking
 the trace over the third component and imposing the constraints
 (\ref{t523}).

\noindent {\bf 2.} Consider a diagonal $N\times N$ block (numbered
$ii$) of the Lax equation. The l.h.s. of the Lax equations is
  \beq\label{t535}
  \begin{array}{c}
  \displaystyle{
 \hbox{l.h.s.}={\dot \mL}^{ii}(z)={\dot p}_i1_N+\tr_2({\dot
 \mS}_2^{ii}R_{12}^{z,(0)}P_{12})\stackrel{(\ref{t305})}{=}{\dot p}_i1_N+\tr_2({\dot
 \mS}_2^{ii}r_{12}(z))\,.
 }
  \end{array}
 \eq
 The r.h.s. of the Lax equation is as follows:
  \beq\label{t536}
  \begin{array}{c}
  \displaystyle{
 \hbox{r.h.s.}=[\mL^{ii},\mM^{ii}]
 +\sum\limits_{k:k\neq i}^M \Big(\mL^{ik}\mM^{ki}-\mM^{ik}\mL^{ki}\Big)\,.
 }
  \end{array}
 \eq
 The commutator term in (\ref{t536}) provides the commutator term in
 the equations of motion
 (\ref{t525}) since it is the input from the internal $ii$-th top's
 dynamics, and this was derived in \cite{LOZ16}. See
 (\ref{c406})-(\ref{c410}). In order to simplify expression in the
 sum we need the following degeneration of (\ref{t003}):
 \beq\label{t71}
  \begin{array}{c}
  \displaystyle{
 R^z_{12}(x)
 R^{z}_{23}(y)=R^{z}_{13}(x+y)r_{12}(x)+r_{23}(y)R^z_{13}(x+y)-\p_z
 R_{13}^z(x+y)\,,
 }
 \end{array}
 \eq
It corresponds to $\hbar=\eta=z$. In the scalar case it is the
identity (\ref{a10}). In the limit $x=q=-y$ from (\ref{t71}) we get
 \beq\label{t72}
  \begin{array}{c}
  \displaystyle{
 R^z_{12}(q)
 R^{z}_{23}(-q)=R^{z,(0)}_{13}r_{12}(q)-r_{32}(q)R^{z,(0)}_{13}-\p_z
 R_{13}^{z,(0)}+F^0_{32}(q)P_{13}\,,
 }
 \end{array}
 \eq
or, using (\ref{t305})
 \beq\label{t73}
  \begin{array}{c}
  \displaystyle{
 R^z_{12}(q)
 R^{z}_{23}(-q)=(r_{13}(z)r_{32}(q)-r_{32}(q)r_{13}(z))P_{13}-F^0_{13}(z)P_{13}+F^0_{32}(q)P_{13}\,.
 }
 \end{array}
 \eq
 By differentiating (\ref{t73}) with respect to $q$ we obtain
 \beq\label{t537}
  \begin{array}{c}
  \displaystyle{
 R^z_{12}(q)
 F^{z}_{23}(-q)-F^z_{12}(q)
 R^{z}_{23}(-q)=[F^0_{32}(q),r_{13}(z)]P_{13}-\p_qF^0_{32}(q)P_{13}\,.
 }
 \end{array}
 \eq
 For $k\neq i$ consider
 \beq\label{t538}
  \begin{array}{c}
  \displaystyle{
 \mL^{ik}\mM^{ki}-\mM^{ik}\mL^{ki}=
 }
 \\ \ \\
   \displaystyle{
=\tr_{23}\Big(\Big(R^z_{12}(q_{ik})
F^{z}_{23}(q_{ki})-F^z_{12}(q_{ik})
 R^{z}_{23}(q_{ki})\Big)P_{12}P_{13}\mS_2^{ik}\mS^{ki}_3\Big)\stackrel{(\ref{t537})}{=}
 }
  \\ \ \\
   \displaystyle{
 =\tr_{23}\Big(\Big( [F^0_{32}(q_{ik}),r_{13}(z)]P_{13}-\p_{q_{i}}F^0_{32}(q_{ik})P_{13}
 \Big)P_{12}P_{13}\mS_2^{ik}\mS^{ki}_3\Big)\,.
 }
 \end{array}
 \eq
The commutator term in the obtained expression yields the sum term
in the equations of motion (\ref{t525}), while the last term in
(\ref{t538}) provides equations of motion (\ref{t526}). Indeed,
 \beq\label{t539}
  \begin{array}{c}
  \displaystyle{
\tr_{23}\Big(\Big( \p_{q_{i}}F^0_{32}(q_{ik})P_{13}
 \Big)P_{12}P_{13}\mS_2^{ik}\mS^{ki}_3\Big)=1_N \tr_{23}\Big( \p_{q_{i}}F^0_{32}(q_{ik})
 P_{23}\mS_2^{ik}\mS^{ki}_3\Big)\,,
 }
 \end{array}
 \eq
and the momenta is the scalar component in the l.h.s. (\ref{t535}).
 $\blacksquare$


\subsection{Hamiltonian description}

\paragraph{The Hamiltonian function.} Let us compute the Hamiltonian for the model
(\ref{t50})-(\ref{t526}). It comes from the generating function
 \beq\label{t550}
  \begin{array}{c}
  \displaystyle{
 \frac{1}{2N}\,\tr(\mL^2(z))=\frac{1}{2N}\sum\limits_{i=1}^M\tr\Big(\mL^{ii}(z)\Big)^2+\frac{1}{2N}\sum\limits_{i\neq
 j}^M \tr\Big(\mL^{ij}(z)\mL^{ji}(z)\Big)\,.
 }
 \end{array}
 \eq
 Consider
 \beq\label{t551}
  \begin{array}{c}
  \displaystyle{
 \tr\Big(\mL^{ii}(z)\Big)^2=Np_i^2+2p_i\,\tr_{12}\Big(r_{12}(z)\mS^{ii}_2\Big)+\tr_{123}\Big(r_{12}(z)r_{13}(z)\mS^{ii}_2\mS^{ii}_3\Big)\,.
 }
 \end{array}
 \eq
 As before, the numbered tensor components are $\Mat$-valued. In
 order to simplify (\ref{t551}) we use the identity (see
 \cite{LOZ9})
 \beq\label{t552}
  \begin{array}{c}
  \displaystyle{
  r_{12}(z)r_{13}(z\!+\!w)-r_{23}(w)r_{12}(z)+r_{13}(z\!+\!w)r_{23}(w)=m_{12}(z)+m_{23}(w)+m_{13}(z\!+\!w)\,,
 }
 \end{array}
 \eq
 which can be treated as a half of the classical Yang-Baxter
 equation\footnote{The difference of two such equations gives the
 classical Yang-Baxter equation for the classical $r$-matrix.}.
 In the limit $w\rightarrow 0$ (\ref{t552}) yields
 \beq\label{t553}
  \begin{array}{c}
  \displaystyle{
  r_{12}(z)r_{13}(z)=r_{23}^{(0)}r_{12}(z)-r_{13}(z)r_{23}^{(0)}-F^0_{13}(z)P_{23}+m_{12}(z)+m_{23}(0)+m_{13}(z)\,.
 }
 \end{array}
 \eq
 Also, we are going to use the following $R$-matrix property:
 \beq\label{t554}
  \begin{array}{c}
  \displaystyle{
 \tr_1R^q_{12}(z)=\tr_2R^q_{12}(z)=\ti\phi(z,q)1_N\,,
 }
 \end{array}
 \eq
 where $\ti\phi(z,q)$ is the Kronecker function (\ref{a01}) but with
 possibly different normalization factor and normalization of
 arguments.
 The property (\ref{t554}) holds true in the elliptic case (\ref{c62}) as well as for its trigonometric and rational degenerations.
 From (\ref{t554}), expansion (\ref{t004}) and (\ref{a11}) we
 also have similar properties for $\tr_1r_{12}(z)={\ti E}_1(z)$ and
 $\tr_1m_{12}(z)$ -- they are scalar operators:
 \beq\label{t5541}
  \begin{array}{c}
  \displaystyle{
 \tr_1R^q_{12}(z)=q^{-1}1_N+\tr_1r_{12}(z)+q\tr_1m_{12}(z)+O(q^2)\,.
 }
 \end{array}
 \eq
  Return now to (\ref{t551}). On the constraints (\ref{t523}) the second
  term is equal to $2p_i{\ti E}_1(z)\hbox{const}$. After summation
  over $i$ it provides the Hamiltonian proportional to $\sum^M_{i=1}
  p_i$.
  Plugging (\ref{t553}) into the last term of (\ref{t551})
  we get
 \beq\label{t555}
  \begin{array}{c}
  \displaystyle{
 \tr_{123}\Big(r_{12}(z)r_{13}(z)\mS^{ii}_2\mS^{ii}_3\Big)=
 }
 \\ \ \\
  \displaystyle{
 =\tr_{123}\Big(\Big(r_{23}^{(0)}r_{12}(z)-r_{13}(z)r_{23}^{(0)}-F^0_{13}(z)P_{23}+
 m_{23}(0)+m_{12}(z)+m_{13}(z)\Big)\mS^{ii}_2\mS^{ii}_3\Big)\,.
 }
 \end{array}
 \eq
 Due to (\ref{t554}) the first two terms are cancelled out after
 taking the trace over the component $1$. By the same reason the last two terms in
 (\ref{t555}) provide
 $2\tr_{1}(m_{12}(z))\tr_{23}(\mS^{ii}_2\mS^{ii}_3)$. These are
 constants on the constraints (\ref{t523}). The rest of the terms
 are
 \beq\label{t556}
  \begin{array}{c}
  \displaystyle{
 \tr_{123}\Big(\Big(-F_{13}(z)P_{23}+
 m_{23}(0)\Big)\mS^{ii}_2\mS^{ii}_3\Big)\stackrel{(\ref{t554})}{=}
 {\ti E}_2(z)\tr\Big(\mS^{ii}\Big)^2+N\tr_{23}\Big(m_{23}(0)\mS^{ii}_2\mS^{ii}_3\Big)\,,
 }
 \end{array}
 \eq
 where ${\ti E}_2(z)1_N=-\tr_1(F^0_{13}(z))=-\p_z\tr_1(r_{13}(z))=-\p_z{\ti E}_1(z)1_N$. It is a scalar
 function coming from (\ref{t5541}) and similar to $E_2(z)$
 (\ref{a04}). The factor $N$ in the last term comes from $\tr_1$. The first term in (\ref{t556}) is a part of the Casimir
 function $\tr\mS^2$, and the second one is
 $H^{\hbox{\tiny{top}}}(\mS^{ii})$ from (\ref{t01}):
 \beq\label{t557}
  \begin{array}{c}
  \displaystyle{
 H^{\hbox{\tiny{top}}}(\mS^{ii})=\frac{1}{2}\,\tr_{12}\Big(m_{12}(0)\mS^{ii}_1\mS^{ii}_2\Big)\,.
 }
 \end{array}
 \eq
Next, consider
 \beq\label{t74}
  \begin{array}{c}
  \displaystyle{
 \tr\Big(\mL^{ij}(z)\mL^{ji}(z)\Big)=\tr_{123}\Big(R^z_{12}(q_{ij})P_{12}
 R^{z}_{13}(q_{ji})P_{13}\,\mS_2^{ij}\mS_3^{ji}\Big)=
 }
 \\ \ \\
  \displaystyle{
 =\tr_{123}\Big(R^z_{12}(q_{ij})
 R^{z}_{23}(q_{ji})P_{12}P_{13}\,\mS_2^{ij}\mS_3^{ji}\Big)\stackrel{(\ref{t73})}{=}
 }
  \\ \ \\
  \displaystyle{
 =\tr_{123}\Big(\Big(
[r_{13}(z),r_{32}(q_{ij})]-F^0_{13}(z)+F^0_{32}(q_{ij})
 \Big)P_{23}\,\mS_2^{ij}\mS_3^{ji}\Big)\,.
 }
 \end{array}
 \eq
 Again, the commutator term vanishes after taking the trace over the first tensor
 component. Therefore,
 \beq\label{t558}
  \begin{array}{c}
  \displaystyle{
 \tr\Big(\mL^{ij}(z)\mL^{ji}(z)\Big)
 =\tr_{123}\Big(\Big(
-F^0_{13}(z)+F^0_{32}(q_{ij})
 \Big)P_{23}\,\mS_2^{ij}\mS_3^{ji}\Big)=
 }
 \\ \ \\
  \displaystyle{
 ={\ti E}_2(z)\tr\Big(\mS^{ij}\mS^{ji}\Big)+N\tr_{12}\Big(
 F^0_{21}(q_{ij}) P_{12}\,\mS_1^{ij}\mS_2^{ji}\Big)\,.
 }
 \end{array}
 \eq
 Finally, for the potential term from (\ref{t01}) we have
 \beq\label{t559}
  \begin{array}{c}
  \displaystyle{
 \mU(\mS^{ij},\mS^{ji},q_{ij})
 =\tr_{12}\Big(
 F^0_{21}(q_{ij}) P_{12}\,\mS_1^{ij}\mS_2^{ji}\Big)
 }
 \end{array}
 \eq
 and \underline{the Hamiltonian (\ref{t01}) is of the form:}
 \beq\label{t560}
 \begin{array}{c}
  \displaystyle{
 \mH=\sum\limits_{i=1}^M \frac{p_i^2}{2}+\frac{1}{2}\sum\limits_{i=1}^M
 \tr_{12}\Big(m_{12}(0)\mS^{ii}_1\mS^{ii}_2\Big)+\sum\limits_{i<j}^M
 \tr_{12}\Big(
 F^0_{21}(q_{ij}) P_{12}\,\mS_1^{ij}\mS_2^{ji}\Big)\,.
 }
 \end{array}
 \eq
In $M=1$ case $\mH$ reproduce the Hamiltonian of the integrable top,
while in the $M=1$ case we obtain the spin Calogero-Moser
Hamiltonian (\ref{t06}) up to terms containing $S_{ii}$ -- they are
constant in this case (\ref{t07}).

\paragraph{Poisson brackets.} The Poisson structure (before reduction
(\ref{t205})) consists of the canonical brackets for positions and
momenta
 \beq\label{t561}
 \begin{array}{c}
  \displaystyle{
  \{p_i,q_j\}=\delta_{ij}\,,\quad \{p_i,p_j\}=\{q_i,q_j\}=0\,,\quad
  i=1...M
 }
 \end{array}
 \eq
and the linear Poisson-Lie brackets for the $\mS$ variables. They
are of the form (\ref{a25}) but for $\MatNM$ case instead of $\MatM$
in (\ref{a25}). It is convenient to write down these brackets in
terms of $\Mat$-valued blocks $\mS^{ij}$. For $i,j,k,l=1...M$ and
$a,b,c,d=1...N$:
 \beq\label{t562}
 \begin{array}{c}
  \displaystyle{
  \{\mS^{ij}_{ab},\mS^{kl}_{cd}\}=\mS^{kj}_{cb}\,\delta^{il}\,\delta_{ad}-\mS^{il}_{ad}\,\delta^{kj}\,\delta_{bc}
 }
 \end{array}
 \eq
 or
 \beq\label{t563}
 \begin{array}{c}
  \displaystyle{
  \{\mS_1^{ij},\mS_2^{kl}\}=P_{12}\,\mS_1^{kj}\,\delta^{il}-\mS_{1}^{il}\,P_{12}\,\delta^{kj}\,,
 }
 \end{array}
 \eq
where $P_{12}$ as before the permutation operator in $\Mat^{\otimes
2}$. For the diagonal blocks we have
 \beq\label{t564}
 \begin{array}{c}
  \displaystyle{
  \{\mS_1^{ii},\mS_2^{kk}\}=[P_{12},\mS_1^{ii}]\,\delta^{ik}\,.
 }
 \end{array}
 \eq
 It is verified directly that
 \begin{predl}
 The Poisson structure (\ref{t561}),
 (\ref{t563}) and the Hamiltonian (\ref{t560}) provides equations of
 (\ref{t524})-(\ref{t526}), i.e. for the l.h.s. of the Lax equation
 (\ref{t3000}) we have
 \beq\label{t565}
 \begin{array}{c}
  \displaystyle{
 {\dot \mL}(z)=\{\mH,\mL(z)\}\,.
 }
 \end{array}
 \eq
 \end{predl}

\subsection{Interacting tops} Suppose the matrix $\mS$ is of rank
one, i.e. (\ref{t201}) is fulfilled.
 Consider the potential
 \beq\label{t77}
  \begin{array}{c}
  \displaystyle{
 \tr_{12}\Big(
 F^0_{21}(q_{ij})
 P_{12}\,\mS_1^{ij}\mS_2^{ji}\Big)=\sum\limits_{a,b,c,d=1}^N (F^0_{12}(q_{ji})
 P_{12})_{ab,cd}\mS^{ij}_{ba}\mS^{ji}_{dc}\,.
 }
 \end{array}
 \eq
The right multiplication of an element
 $T_{12}=\sum\limits_{i,j,k,l=1}^N
T_{ijkl}E_{ij}\otimes
 E_{kl}\in\Mat^{\otimes 2}$  by permutation operator $P_{12}$  yields
$T_{ijkl}\rightarrow T_{ilkj}$, i.e.
 \beq\label{t78}
  \begin{array}{c}
  \displaystyle{
 \tr_{23}\Big(
 F^0_{32}(q_{ij})
 P_{23}\,\mS_2^{ij}\mS_3^{ji}\Big)=\sum\limits_{a,b,c,d=1}^N (F^0_{12}(q_{ji})
 )_{ad,cb}\mS^{ij}_{ba}\mS^{ji}_{dc}\,.
 }
 \end{array}
 \eq
In the rank 1 case we have
 \beq\label{t79}
  \begin{array}{c}
  \displaystyle{
\mS^{ij}_{ba}\mS^{ji}_{dc}=\xi^i_b\eta^j_a\xi^j_d\eta^i_c=\mS^{ii}_{bc}\mS^{jj}_{da}\,.
 }
 \end{array}
 \eq
 Therefore,
 \beq\label{t80}
  \begin{array}{c}
  \displaystyle{
 \tr_{23}\Big(
 F^0_{32}(q_{ij})
 P_{23}\,\mS_2^{ij}\mS_3^{ji}\Big)=\tr_{12}\Big(F^0_{12}(q_{ji})\mS^{jj}_1\mS^{ii}_2\Big)=
 \tr_{12}\Big(F^0_{12}(q_{ij})\mS^{ii}_1\mS^{jj}_2\Big)\,.
 }
 \end{array}
 \eq
 \underline{The Hamiltonian of interacting tops model} acquires the form:
 \beq\label{t81}
  \begin{array}{c}
  \displaystyle{
\mH^{\hbox{\tiny{tops}}}=\sum\limits_{i=1}^M
\frac{p_i^2}{2}+\frac{1}{2}\sum\limits_{i=1}^M
 \tr_{12}\Big(m_{12}(0)\mS^{ii}_1\mS^{ii}_2\Big)+\sum\limits_{i<j}^M
 \tr_{12}\Big(F^0_{12}(q_{ij})\mS^{ii}_1\mS^{jj}_2\Big)\,.
 }
 \end{array}
 \eq
 From the Poisson brackets (\ref{t561}), (\ref{t564}) we get the corresponding \underline{equations of
 motion:}
  \beq\label{t570}
  \begin{array}{c}
  \displaystyle{
 {\dot \mS}^{ii}=[\mS^{ii},\tr_2(m_{12}(0)\mS^{ii}_2)]+\sum\limits_{k:k\neq
 i}^M\, [\mS^{ii},\tr_2(F^0_{12}(q_{ik})\mS^{kk}_2)]\,,
 }
  \end{array}
 \eq
  \beq\label{t571}
  \begin{array}{c}
  \displaystyle{
 {\dot p}_i=-\sum\limits_{k:k\neq
 i}^M\tr_{12}\Big(\p_{q_i}F^0_{12}(q_{ik})\mS^{ii}_1\mS^{kk}_2\Big)\,.
 }
  \end{array}
 \eq
 In this model we are left with $M$ matrix variables $\mS^{ii}\in\Mat$ of rank one. It is notable
 that the spin part of the phase space (\ref{t206}) is isomorphic to
 a product of $M$ minimal coadjoint orbits (\ref{t0091}):
  \beq\label{t572}
  \begin{array}{c}
  \displaystyle{
  \mathcal O_{NM}^{\hbox{\tiny{min}}}//{\mathfrak
  H}_{NM}'\cong\underbrace{ O_{N}^{\hbox{\tiny{min}}}\times...\times O_{N}^{\hbox{\tiny{min}}}
  }_{M\
  times}\,.
 }
 \end{array}
 \eq
 Notice that the orbits $O_{N}^{\hbox{\tiny{min}}}$ come from the
 constraints conditions (\ref{t523}).
 Hence it appears that

 {\bf 1.} For the model of interacting tops the constraints (\ref{t523})
 play the role of fixation of the Casimir functions for $M$ copies
 of  ${\rm gl}_N^*$ (of rank one). Consequently, equations of motion
 (\ref{t570}) are not changed after reduction. For the $N=1$ case (the spin Calogero-Moser model)
 we get ${\dot S}_{ii}=0$ since the r.h.s. of (\ref{t570}) consists
 of commutators.

 {\bf 2.} The model of interacting tops is formulated in terms of $M$
 $\Mat$-valued variables of rank one, describing the minimal
 coadjoint orbits. The integrability condition is that all Casimir
 functions $\tr(\mS^{ii})$ are equal to each other\footnote{More precisely,
we can not confirm that the model is not integrable in the case
$\tr(\mS^{ii})\neq \tr(\mS^{jj})$, but the presented Lax pair does
not work in this case.}.

 {\bf 3.} The spin part of the phase space for the model of
 interacting tops coincides with the phase space of ${\rm GL}_N$
 classical spin chain on $M$ sites with the spins described by
 minimal coadjoint orbits at each site.

Let us also remark that the top like models with matrix-valued
variables were studied in \cite{LOZ16,Z2} and \cite{Rubt}. In
contrast to these papers here we deal with the models, where the
matrix variables have their own internal dynamics.


\section{Classical $r$-matrix}
\setcounter{equation}{0}

In this Section we describe the classical $r$-matrix structure for
the Lax matrix (\ref{t51}). Since $\mL\in\MatNM$ then the
corresponding  classical ${\rm gl}_{NM}$ $r$-matrix ${\bf
r}\in\MatNM^{\otimes 2}$.
 Recall that for the Lax matrix we use the matrix basis (\ref{t50}), in which
$\mL\in\MatM\otimes \Mat$. Let the $\MatM$-valued tensor components
be numbered by primed numbers, and the $\Mat$-valued components --
without primes (as before). Introduce the following $r$-matrix:
  \beq\label{t575}
  \begin{array}{c}
  \displaystyle{
  {\bf r}_{1'2'12}(z,w)=\sum\limits_{i=1}^M\stackrel{1'}{E}_{ii}\otimes\stackrel{2'}{E}_{ii}
  \otimes\, r_{12}(z-w)
+\sum\limits_{i\neq j}^M
\stackrel{1'}{E}_{ij}\otimes\stackrel{2'}{E}_{ji}\otimes
R_{12}^{z-w}(q_{ij})P_{12}\,,
 }
 \end{array}
 \eq
 so that ${\bf r}_{1'2'12}\in\MatM^{\otimes 2}\otimes\Mat^{\otimes
 2}$. In the case $M=1$ we come to a non-dynamical $r$-matrix describing the top model,
 while in the $N=1$ we reproduce the dynamical $r$-matrix of the spin Calogero-Moser model (\ref{a21}).
  $r$-matrices of these type are known in ${\rm gl}_{NM}$ case and can be extended for arbitrary complex semisimple Lie algebras
 \cite{ESch,Feher,LOSZ3}.
 In the elliptic case (\ref{t575}) is known in the quantum
 case as well \cite{LOSZ4}. At the same time (\ref{t575}) includes
 the cases, which have not been described yet. For instance, the new
 cases correspond to the rational $R_{12}^z(q)$-matrix from
 \cite{LOZ8}.
 Similarly to the Lax equations the construction of the $r$-matrix (\ref{t575}) is based on the associative
 Yang-Baxter equation (\ref{t003}) and its degenerations.


\begin{predl}
Consider an $R$-matrix satisfying the associative Yang-Baxter
 equation (\ref{t003}), the classical limit (\ref{t004}) and the set
 of properties from the Section \ref{S21}.
 Then for the Lax pair (\ref{t50})-(\ref{t51}) the following
 classical exchange relation holds true:
  \beq\label{t576}
  \begin{array}{c}
  \displaystyle{
 \{\mL_{1'1}(z),\mL_{2'2}(w)\}=[\mL_{1'1}(z),{\bf r}_{1'2'12}(z,w)]-[\mL_{2'2}(w),{\bf
 r}_{2'1'21}(w,z)]-
 }
 \\ \ \\
  \displaystyle{
 -\sum\limits_{k=1}^M\tr(\mS^{kk})\p_{q_k}{\bf r}_{1'2'12}(z,w)\,,
 }
 \end{array}
 \eq
where
  \beq\label{t577}
  \begin{array}{c}
  \displaystyle{
 \mL_{1'1}(z)=\sum\limits_{i,j=1}^ME_{ij}\otimes 1_M\otimes \mL^{ij}(z)\otimes
 1_N\,,
 }
 \end{array}
 \eq
  \beq\label{t5771}
  \begin{array}{c}
  \displaystyle{
 \mL_{2'2}(w)=\sum\limits_{k,l=1}^M 1_M\otimes E_{kl}\otimes
 1_N\otimes  \mL^{kl}(w)\,.
 }
 \end{array}
 \eq
The Poisson brackets in the l.h.s. of (\ref{t576}) are given by
(\ref{t561})-(\ref{t564}).
\end{predl}
\noindent\underline{\em{Proof:}}\quad
 The proof is direct. Let us demonstrate how to verify (\ref{t576})
 for several components of
 $\stackrel{1'}{E}_{ij}\otimes\stackrel{2'}{E}_{kl}$, which are
 similar to those considered in (\ref{a51})-(\ref{a56}) for the spin
 Calogero-Moser model.

 \underline{the tensor component
 $\stackrel{1'}{E}_{ij}\otimes\stackrel{2'}{E}_{jk}$ ($i\neq j$, $j\neq k$, $i\neq
 k$):}

\noindent l.h.s. of (\ref{t576}):
  \beq\label{t578}
  \begin{array}{c}
  \displaystyle{
 \tr_{34}\Big(R_{13}^z(q_{ij})P_{13}R_{24}^w(q_{jk})P_{24}\{\mS_3^{ij},\mS_4^{jk}\}\Big)=
 -\tr_3\Big(\mS^{ik}_3R_{23}^w(q_{jk})P_{23}R_{13}^z(q_{ij})P_{13}\Big)\,.
 }
 \end{array}
 \eq
\noindent r.h.s. of (\ref{t576}):
  \beq\label{t579}
  \begin{array}{c}
  \displaystyle{
 \tr_3\Big(\mS^{ik}_3( R_{13}^z(q_{ik})P_{13}R_{12}^{z-w}(q_{kj})P_{12}-R_{12}^{z-w}(q_{ij})P_{12}R_{23}^w(q_{ik})P_{23} )\Big)\,.
 }
 \end{array}
 \eq
Expressions (\ref{t578}) and (\ref{t579}) coincide due to
(\ref{t003}) and (\ref{t301}).

 \underline{the tensor component
 $\stackrel{1'}{E}_{ii}\otimes\stackrel{2'}{E}_{ij}$ ($i\neq j$):}

\noindent l.h.s. of (\ref{t576}):
  \beq\label{t580}
  \begin{array}{c}
  \displaystyle{
  \tr_{4}\Big(\{p_i,R_{24}^w(q_{ij})\}P_{24}\,\mS_4^{ij}\Big)+
 \tr_{34}\Big(r_{13}(z)R_{24}^w(q_{ij})P_{24}\{\mS_3^{ii},\mS_4^{ij}\}\Big)=
 }
 \\ \ \\
  \displaystyle{
  =\tr_{3}\Big(\mS_3^{ij}\p_{q_i}R_{23}^w(q_{ij}) P_{23}\Big)-
 \tr_{3}\Big(\mS_3^{ij}R_{23}^w(q_{ij})P_{23}r_{13}(z)\Big)\,.
 }
 \end{array}
 \eq
\noindent r.h.s. of (\ref{t576}):
  \beq\label{t581}
  \begin{array}{c}
  \displaystyle{
 \tr_{3}\Big(\mS_3^{ij}( R_{13}^w(q_{ij})P_{13}R_{12}^{z-w}(q_{ji})P_{12}-r_{12}(z-w)R^w_{23}(q_{ij})P_{23} )\Big)=
 }
 \end{array}
 \eq
Expressions (\ref{t580}) and (\ref{t581}) coincide due to
(\ref{t71}) rewritten through the Fourier symmetry (\ref{t303}) as
  \beq\label{t582}
  \begin{array}{c}
  \displaystyle{
R_{ac}^{q_{ij}}(z) R_{bc}^{q_{ji}}(w) =
        - R_{ab}^{q_{ij}}(z - w) r_{ac}(z) +
            r_{bc}(w) R_{ab}^{q_{ij}}(z - w) -
                \p_{q_i} R_{ab}^{q_{ij}}(z - w)
 }
 \end{array}
 \eq
for distinct $a,b,c$.

 \underline{the tensor component
 $\stackrel{1'}{E}_{ij}\otimes\stackrel{2'}{E}_{ji}$ ($i\neq j$):}

\noindent l.h.s. of (\ref{t576}):
  \beq\label{t583}
  \begin{array}{c}
  \displaystyle{
 \tr_3\Big( \mS_3^{jj}R_{13}^z(q_{ij})P_{13}R_{23}^w(q_{ji})P_{23}-\mS_3^{ii}R_{23}^w(q_{ji})P_{23}R_{13}^z(q_{ij})P_{13}
 \Big)\,.
 }
 \end{array}
 \eq
\noindent r.h.s. of (\ref{t576}):
  \beq\label{t584}
  \begin{array}{c}
  \displaystyle{
 \tr_3\Big( \mS_3^{ii}( r_{13}(z)R_{12}^{z-w}(q_{ij})P_{12}-R_{12}^{z-w}(q_{ij})P_{12}r_{23}(w)
 )
 \Big)+
 }
 \\ \ \\
   \displaystyle{
 +\tr_3\Big( \mS_3^{jj}( -R_{12}^{z-w}(q_{ij})P_{12}r_{13}(z)+r_{23}(w)R_{12}^{z-w}(q_{ij})P_{12}
 )
 \Big)-
 }
  \\ \ \\
   \displaystyle{
 -\tr_3\Big( \Big(\mS_3^{ii}-\mS_3^{jj}\Big)
 \p_{q_i}R_{12}^{z-w}(q_{ij})P_{12}
 \Big)\,.
 }
 \end{array}
 \eq
The last term comes from the second line of (\ref{t576}). Again,
expressions (\ref{t583}) and (\ref{t584}) coincide due to
(\ref{t582}).

The rest of the components are verified similarly. $\blacksquare$


\section{Examples}\label{ex}
\setcounter{equation}{0}

\subsection{Elliptic models}

Let us begin with the elliptic model \cite{ZL,LOSZ4,GrZ}. The Lax
pair is of the form:
  \beq\label{t20}
  \begin{array}{c}
  \displaystyle{
 \mL(z)=\sum\limits_{i,j=1}^M E_{ij}\otimes \mL^{ij}(z)\,,\quad\quad
 \mL^{ij}(z)\in{\rm Mat}_{N}\quad\quad \mL(z)\in{\rm Mat}_{NM}\,,
 }
  \end{array}
 \eq
where
  \beq\label{t21}
  \begin{array}{c}
  \displaystyle{
 \mL^{ij}(z)=\delta_{ij}\Big(p_i1_N+\mS^{\,ii}_{(0,0)}\, 1_N\,E_1(z)+\!\sum\limits_{\al\neq0}\mS^{\,ii}_\al\,
 T_\al\,\vf_\al(z,\om_\al)\Big)+
 }
  \\ \ \\
  \displaystyle{
  +(1-\delta_{ij})\sum\limits_{\al}\mS^{\,ij}_\al\,
  T_\al\,\vf_\al(z,\om_\al+\frac{q_{ij}}{N})\,,
 }
  \end{array}
 \eq
 where the basis (\ref{e904}) in $\Mat$ is used.
Similarly, the $M$-matrix is of the form
  \beq\label{t22}
  \begin{array}{c}
  \displaystyle{
 \mM^{ij}(z)=\delta_{ij}\,\mS^{\,ii}_{(0,0)}\frac{E_1^2(z)-\wp(z)}{2N}\,1_N+
 \frac1N\,\delta_{ij}\sum\limits_{\al\neq0}\mS^{\,ii}_\al\,
T_\al\,f_\al(z,\om_\al)+
  }
  \\ \ \\
  \displaystyle{
  +\frac1N\,(1-\delta_{ij})\sum\limits_{\al}\mS^{\,ij}_\al\,
  T_\al\,f_\al(z,\om_\al+\frac{q_{ij}}{N})\,.
 }
  \end{array}
 \eq
These formulae can be obtained from (\ref{t50})-(\ref{t52}) and the
$R$-matrix (\ref{c62}) together with (\ref{a08}).

The Lax equations hold on the constraints
  \beq\label{t23}
  \begin{array}{c}
  \displaystyle{
\mS^{\,ii}_{(0,0)}=\hbox{const}\,,\quad\forall i\,.
 }
  \end{array}
 \eq
 Instead of the standard basis (\ref{t03}) here we use the basis (\ref{e904}) for each $N\times N$ block.
  Then the Poisson structure (\ref{t05}) takes the form
  \beq\label{t240}
  \begin{array}{c}
  \displaystyle{
 \{\mS^{\,ij}_\al,\mS^{\,kl}_\be\}=\delta_{il}\,\kappa_{\al,\be}\,\mS^{\,kj}_{\al+\be}-
 \delta_{kj}\,\kappa_{\be,\al}\,\mS^{\,il}_{\al+\be}\,,
 }
  \end{array}
 \eq
 where $\kappa_{\al,\be}$ are the constants from (\ref{e905}).

 The Hamiltonian easily follows from $\frac1{2N}\tr\mL^2(z)=\frac1{2N}E_2(z)\tr(\mS^2)+\mH$ due to (\ref{e906}) and (\ref{a08}):
  \beq\label{t24}
  \begin{array}{c}
  \displaystyle{
 \mH=\frac{1}{2}\sum\limits_{i=1}^M
 p_i^2-\frac{1}{2}\sum\limits_{i=1}^M\sum\limits_{\al\neq 0}
 \mS^{\,ii}_\al\, \mS ^{ii}_{-\al}\,E_2(\om_\al)-
 \frac{1}{2}\sum\limits_{i\neq j}^M\sum\limits_{\al}
 \mS^{\,ij}_{\al}\,\mS ^{ji}_{-\al}\,E_2(\om_\al+\frac{q_{ij}}{N})\,.
 }
  \end{array}
 \eq
 Let us show how this Hamiltonian is reproduced from the general
 formula
(\ref{t560}). In order to get the second term in (\ref{t24}) one
should substitute $m_{12}(0)$ into (\ref{t560}) from (\ref{c64}) and
use relation (\ref{a13}). For evaluation of the last sum in
(\ref{t560}) we need to calculate $F^0_{12}(q)P_{12}$. The answer
for $F_{12}^0(q)$ is given in (\ref{c65}). Multiply it by
$P_{12}=(1/N)\sum_bT_b\otimes T_{-b}$ from the left:
  \beq\label{t92401}
  \begin{array}{c}
  \displaystyle{
 F^0_{12}(q)P_{12}=-\frac{1}{N^2}\,E_2(q)\sum\limits_{b}T_b\otimes T_{-b}
 }
 \\ \ \\
  \displaystyle{
 +\frac{1}{N^2}\sum\limits_{a\neq(0,0);b} \vf_a(q,\om_a)(E_1(q+\om_a)-E_1(q)+2\pi\imath\p_\tau\om_\al)\,
 \kappa_{a,b}^2\, T_{a+b}\otimes
 T_{-a-b}\,.
 }
  \end{array}
 \eq
 Let us redefine the summation index $b\rightarrow b-a$ in the last
 sum. Since $\kappa_{a,b}=\kappa_{a,b-a}$ we have
  \beq\label{t92402}
  \begin{array}{c}
  \displaystyle{
 F^0_{12}(q)P_{12}=
 }
 \\ \ \\
  \displaystyle{
 \frac{1}{N^2}\sum\limits_{b}T_b\otimes T_{-b}\Big(-E_2(q)
 +\!\sum\limits_{a\neq0;b} \vf_a(q,\om_a)(E_1(q+\om_a)-E_1(q)+2\pi\imath\p_\tau\om_\al)\,
 \kappa_{a,b}^2\Big)\stackrel{(\ref{c69})}{=}
 }
  \\ \ \\
  \displaystyle{
 =- \frac{1}{N^2}\sum\limits_{b}T_b\otimes T_{-b}\,
 E_2(\om_b+\frac{q}{N})\,.
 }
  \end{array}
 \eq
Finally,
  \beq\label{t92403}
  \begin{array}{c}
  \displaystyle{
\tr_{12}\Big(
 F^0_{21}(q_{ij}) P_{12}\,\mS_1^{ij}\mS_2^{ji}\Big)=\sum\limits_{\al}
 \mS^{\,ij}_{\al}\,\mS
 ^{ji}_{-\al}\,E_2(\om_\al+\frac{q_{ij}}{N})\,.
  }
  \end{array}
 \eq

\underline{In the rank 1 case} the answer for the Hamiltonian is
given by (\ref{t81}). Plugging (\ref{c65}) into  (\ref{t81}) we get
  \beq\label{t92404}
  \begin{array}{c}
  \displaystyle{
 \mH^{\hbox{\tiny{tops}}}=\frac{1}{2}\sum\limits_{i=1}^M
 p_i^2-\frac{1}{2}\sum\limits_{i=1}^M\sum\limits_{\al\neq 0}
 \mS^{\,ii}_\al\, \mS ^{ii}_{-\al}\,E_2(\om_\al)-
 }
 \\ \ \\
   \displaystyle{
 -
 \frac{N}{2}\sum\limits_{i\neq j}^M
 \Big( E_2(q_{ij})\mS^{ii}_0\mS^{jj}_0
 -\sum\limits_{a\neq0}
  \vf_a(q_{ij},\om_a)(E_1(q_{ij}+\om_a)-E_1(q_{ij})+2\pi\imath\p_\tau\om_\al)\mS^{ii}_{-\al}\mS^{jj}_\al \Big)\,.
 }
  \end{array}
 \eq
 Let us show how the latter expression appears from (\ref{t24}).
In the rank one case using (\ref{e906}) (so that
$\mS^{ij}_\al=\tr(\mS^{ij}T_{-\al})/N$) we get
  \beq\label{t241}
  \begin{array}{c}
  \displaystyle{
 \mS^{\,ij}_{\al}\,\mS ^{ji}_{-\al}=\frac{\tr(\eta^j\, T_{-\al}\,\xi^i)\,\tr(\eta^i\,
 T_{\al}\,\xi^j)}{{ N}^2}=\frac{\tr(\eta^j\, T_{-\al}\,\xi^i\,\eta^i\,
 T_{\al}\,\xi^j)}{{ N}^2}=\frac{\tr(\mS^{ii}\,T_{\al}\,\mS^{jj}\,T_{-\al})}{{ N}^2}\,.
 }
  \end{array}
 \eq
 In this way the Hamiltonian (\ref{t24}) acquires the form
  \beq\label{t25}
  \begin{array}{c}
  \displaystyle{
 \mH^{\hbox{\tiny{tops}}}=\frac{1}{2}\sum\limits_{i=1}^M
 p_i^2-\frac{1}{2}\sum\limits_{i=1}^M\sum\limits_{\al\neq 0}
 \mS^{ii}_\al\, \mS^{ii}_{-\al}E_2(\om_\al)-
  }
  \\ \ \\
  \displaystyle{
  -
 \frac{1}{2}\sum\limits_{i\neq j}^M\sum\limits_{\al}
 \frac{\tr(\mS^{ii}\,T_{\al}\,\mS^{jj}\,T_{-\al})}{N^2}\,E_2(\om_\al+\frac{q_{ij}}{N})\,,
 }
  \end{array}
 \eq
which is the model of interacting tops of (\ref{t203}) type.
 The last terms in (\ref{t25}) can be simplified in the following
 way. Substitute $\mS^{ii}=\sum_\ga\mS^{ii}_\ga T_\ga$ and $\mS^{jj}=\sum_\ga\mS^{jj}_\mu
 T_\mu$ into (\ref{t25}). It follows from (\ref{e905})-(\ref{e906})
 that
 \beq\label{t251}
  \begin{array}{c}
  \displaystyle{
  \tr(T_\ga T_\al T_\mu
 T_{-\al})=N\kappa^2_{\al,\mu}\delta_{\mu+\ga}\,.
 }
  \end{array}
 \eq
  Therefore,
  \beq\label{t26}
  \begin{array}{c}
  \displaystyle{
 \sum\limits_{\al}
 \frac{\tr(\mS^{ii}\,T_{\al}\,\mS^{jj}\,T_{-\al})}{N^2}\,E_2(\om_\al+\frac{q_{ij}}{N})
 =\frac1N\sum\limits_{\al,\mu}\mS^{ii}_{-\mu}\mS^{jj}_{\mu}E_2(\om_\al+\frac{q_{ij}}{N})\kappa^2_{\al,\mu}\,.
 }
  \end{array}
 \eq
 Using (\ref{c67})-(\ref{c68}) and summing up over $\al$ we obtain
 the last term in (\ref{t92404}).


\subsection{Trigonometric models}
The general classification of the unitary trigonometric $R$-matrices
satisfying associative Yang-Baxter equation was given in \cite{Sch}.
It includes the 7-vertex deformation \cite{Cherednik2} of the
6-vertex $R$-matrix and its ${\rm GL}_N$ generalizations such as the
non-standard $R$-matrix \cite{AHZ}. The integrable tops and related
structures based on these $R$-matrices were described in \cite{KrZ}.

Here we restrict ourselves to the case $N=2$. The 7-vertex
$R$-matrix is of the following form:
  \beq\label{t820}
   \begin{array}{c}
  R_{12}^\hbar(z)=\left(\begin{array}{cccc} \coth(z)+\coth(\hbar) & 0 & 0 & 0\vphantom{\Big|}
  \\ 0 & \sinh^{-1}(\hbar) & \sinh^{-1}(z) & 0\vphantom{\Big|}
  \\ 0 & \sinh^{-1}(z) & \sinh^{-1}(\hbar) & 0\vphantom{\Big|}
  \\ C\sinh(z+\hbar) & 0 & 0 & \coth(z)+\coth(\hbar)          \vphantom{\Big|} \end{array} \right)
  \end{array}
  \eq
  where $C$ is a constant. In the limit $C\rightarrow 0$ the lower left-hand corner
  vanishes and we get the 6-vertex XXZ $R$-matrix.
 For the classical $r$-matrix and its derivative ($F^0_{12}(z)=\p_z r_{12}(z)$) we have
  \beq\label{t821}
   \begin{array}{c}
  r_{12}(z)=\left(\begin{array}{cccc} \coth(z)&0&0&0\\0&0&\sinh^{-1}(z)&0
  \\0&\sinh^{-1}(z)&0&0\\C\sinh(z)&0&0&\coth(z) \end{array} \right)
  \end{array}
  \eq
  and
    \beq\label{t822}
   \begin{array}{c}
  F^0_{12}(q)=\left(\begin{array}{cccc} \displaystyle{-\frac{1}{\sinh^2(q)}} &0&0&0\\0&0&\displaystyle{-\frac{\cosh(q)}{\sinh^2(q)}}&0
  \\0&\displaystyle{-\frac{\cosh(q)}{\sinh^2(q)}}&0&0\\C\cosh(q)&0&0&\displaystyle{-\frac{1}{\sinh^2(q)}} \end{array} \right)
  \end{array}
  \eq
  respectively. The Fourier transformed $F^0$ matrix is of the form:
   \beq\label{t823}
   \begin{array}{c}
  F^0_{12}(q)P_{12}
  =\left(\begin{array}{cccc} \displaystyle{-\frac{1}{\sinh^2(q)}} &0&0&0\\0&\displaystyle{-\frac{\cosh(q)}{\sinh^2(q)}}&0&0
  \\0&0&\displaystyle{-\frac{\cosh(q)}{\sinh^2(q)}}&0\\C\cosh(q)&0&0&\displaystyle{-\frac{1}{\sinh^2(q)}} \end{array} \right)
  \end{array}
  \eq
From the latter matrix using (\ref{t208}) we obtain
  \beq\label{t824}
  \begin{array}{c}
      \displaystyle{
 \mU(\mS^{ij},\mS^{ji},q_i-q_j)=\tr_{12}\Big(
 \p_{q_i}r_{21}(q_{ij}) P_{12}\,\mS_1^{ij}\mS_2^{ji}\Big)=
  }
  \\ \ \\
        \displaystyle{
 =-\frac{1}{\sinh^2(q_{ij})} \Big(\mS^{ij}_{11}\mS^{ji}_{11}+\mS^{ij}_{22}\mS^{ji}_{22}\Big)
 -\frac{\cosh(q_{ij})}{\sinh^2(q_{ij})}\Big(\mS^{ij}_{11}\mS^{ji}_{22}+\mS^{ij}_{22}\mS^{ji}_{11}\Big)
 +C\cosh(q_{ij})\mS^{ij}_{12}\mS^{ji}_{12}\,.
  }
  \end{array}
  \eq
 Similarly, using (\ref{t209}) and (\ref{t822}) we get the potential
 for the model of interacting tops:
  \beq\label{t825}
  \begin{array}{c}
      \displaystyle{
 \mV(\mS^{ii},\mS^{jj},q_i-q_j)
 =\tr_{12}\Big(\p_{q_i}r_{12}(q_{ij})\mS^{ii}_1\mS^{jj}_2\Big)=
 }
   \\ \ \\
        \displaystyle{
 =-\frac{1}{\sinh^2(q_{ij})} \Big(\mS^{ii}_{11}\mS^{jj}_{11}+\mS^{ii}_{22}\mS^{jj}_{22}\Big)
 -\frac{\cosh(q_{ij})}{\sinh^2(q_{ij})}\Big(\mS^{ii}_{12}\mS^{jj}_{21}+\mS^{ii}_{21}\mS^{jj}_{12}\Big)
 +C\cosh(q_{ij})\mS^{ii}_{12}\mS^{jj}_{12}\,.
  }
  \end{array}
  \eq
The top Hamiltonian $\mH^{\hbox{\tiny{top}}}(\mS^{ii})$ entering
(\ref{t01}) or (\ref{t203}) is of the form:
  \beq\label{t8251}
  \begin{array}{c}
      \displaystyle{
 \mH^{\hbox{\tiny{top}}}(\mS^{ii})=\frac12\Big((\mS^{ii}_{11})^2 +(\mS^{ii}_{22})^2
 \Big)+C(\mS^{ii}_{12})^2\,.
  }
  \end{array}
  \eq

\subsection{Rational models}
 The rational $R$-matrices satisfying the required properties are
 represented by the 11-vertex deformation \cite{Cherednik2} of the 6-vertex XXX (Yang's)
 $R$-matrix. Its higher rank analogues were derived in
 \cite{Smirnov} and \cite{LOZ8}. As in trigonometric case here we restrict ourselves to the case $N=2$. The 11-vertex
$R$-matrix is of the following form:
 \beq\label{t830}
 \begin{array}{c}
  \displaystyle{R_{12}^\hbar(z)=
 \left( \begin{array}{cccc} {\hbar}^{-1}+{z}^{-1}&0&0&0
\\\noalign{\medskip}-\hbar-z&{\hbar}^{-1}&{z}^{-1}&0\\\noalign{\medskip}
-\hbar-z&{z}^{-1}&{\hbar}^{-1}&0\\\noalign{\medskip}-{\hbar}^{3}-2\,z{\hbar}^{2}-2\,\hbar
\,{z}^{2}-{z}^{3}&\hbar+z&\hbar+z&{\hbar}^{-1}+{z}^{-1}
\end{array} \right)
  }
 \end{array}
 \eq
In order to get the XXX $R$-matrix one may take the limit
$\lim\limits_{\epsilon\rightarrow
0}\epsilon^{-1}R^{\epsilon\hbar}(\epsilon z)$.

The classical $r$-matrix, the $F^0_{12}$ matrix and its Fourier dual
are of the form:
 \beq\label{t831}
 \begin{array}{c}
  \displaystyle{r_{12}(z)=
 \left( \begin{array}{cccc} {z}^{-1}&0&0&0
\\\noalign{\medskip}-z&0&{z}^{-1}&0\\\noalign{\medskip}
-z&{z}^{-1}&0&0\\\noalign{\medskip} -{z}^{3}&z&z&{z}^{-1}
\end{array} \right)
  }
 \end{array}
 \eq
 \beq\label{t832}
 \begin{array}{c}
  \displaystyle{F^0_{12}(q)=
 \left( \begin{array}{cccc} -{q}^{-2}&0&0&0
\\\noalign{\medskip}-1&0&-{q}^{-2}&0\\\noalign{\medskip}
-1&-{q}^{-2}&0&0\\\noalign{\medskip} -3{q}^{2}&1&1&-{q}^{-2}
\end{array} \right)
  }
 \end{array}
 \eq
 \beq\label{t833}
 \begin{array}{c}
  \displaystyle{F^0_{12}(q)P_{12}=
 \left( \begin{array}{cccc} -{q}^{-2}&0&0&0
\\\noalign{\medskip}-1&-{q}^{-2}&0&0\\\noalign{\medskip}
-1&0&-{q}^{-2}&0\\\noalign{\medskip} -3{q}^{2}&1&1&-{q}^{-2}
\end{array} \right)
  }
 \end{array}
 \eq
 From (\ref{t833}) using (\ref{t208}) we obtain
  \beq\label{t834}
  \begin{array}{c}
      \displaystyle{
 \mU(\mS^{ij},\mS^{ji},q_i-q_j)=-\frac{1}{(q_i-q_j)^2}
 \Big(\mS^{ij}_{11}\mS^{ji}_{11}+\mS^{ij}_{22}\mS^{ji}_{22}+\mS^{ij}_{11}\mS^{ji}_{22}+\mS^{ij}_{22}\mS^{ji}_{11}\Big)+
  }
  \\ \ \\
        \displaystyle{
 +\mS^{ij}_{12}\mS^{ji}_{22}+\mS^{ij}_{22}\mS^{ji}_{12}-\mS^{ij}_{12}\mS^{ji}_{11}-\mS^{ij}_{11}\mS^{ji}_{12}
 -3(q_i-q_j)^2\mS^{ij}_{12}\mS^{ji}_{12}\,.
  }
  \end{array}
  \eq
 Similarly, from (\ref{t832}) using (\ref{t209}) we obtain
  \beq\label{t835}
  \begin{array}{c}
      \displaystyle{
 \mV(\mS^{ii},\mS^{jj},q_i-q_j)=-\frac{1}{(q_i-q_j)^2}
 \Big(\mS^{ii}_{11}\mS^{jj}_{11}+\mS^{ii}_{22}\mS^{jj}_{22}+\mS^{ii}_{12}\mS^{jj}_{21}+\mS^{ii}_{21}\mS^{jj}_{12}\Big)+
  }
  \\ \ \\
        \displaystyle{
 +\mS^{ii}_{12}\mS^{jj}_{22}+\mS^{ii}_{22}\mS^{jj}_{12}-\mS^{ii}_{12}\mS^{jj}_{11}-\mS^{ii}_{11}\mS^{jj}_{12}
 -3(q_i-q_j)^2\mS^{ii}_{12}\mS^{jj}_{12}\,.
  }
  \end{array}
  \eq
The top Hamiltonian $\mH^{\hbox{\tiny{top}}}(\mS^{ii})$ entering
(\ref{t01}) or (\ref{t203}) is of the form:
  \beq\label{t836}
  \begin{array}{c}
      \displaystyle{
 \mH^{\hbox{\tiny{top}}}(\mS^{ii})=\mS^{ii}_{12}(\mS^{ii}_{22}-\mS^{ii}_{11})\,.
  }
  \end{array}
  \eq
%


\section{Discussion}
\setcounter{equation}{0}

A possible application of the obtained family of integrable models
is
 in constructing quantum integrable anisotropic long-rang
spin chains. The basic idea is that such spin chains appear from the
models of interacting tops by the so-called freezing trick likewise
the Haldane-Shastry-Inozemtsev spin chains \cite{HS} come from the
ordinary spin Calogero-Moser-Sutherland models. A direct
quantization of the interacting tops is a separate problem, which
will be discussed elsewhere. At the same time the quantum
Hamiltonian of interacting tops appears in the so-called
 $R$-matrix-valued Lax pairs for the
(classical) spinless Calogero-Moser model \cite{LOZ9,SeZ,GrZ}. These
are the Lax pairs in a large space $\MatM\otimes\Mat^{\otimes M}$:
  \beq\label{x51}
  \begin{array}{c}
  \displaystyle{
 {\mathcal L}^{\hbox{\tiny{CM}}}=\sum\limits_{a,b=1}^{M} E_{ab}\otimes \mathcal L_{ab}\,,
\ \ \ \mathcal L_{ab}=\delta_{ab}p_a\,1_N^{\otimes
M}+\nu(1-\delta_{ab})R_{ab}^{z}\,,\ \ \
R_{ab}^{z}=R_{ab}^{z}(q_a-q_b)\,.
 }
 \end{array}
 \eq
and similarly for the accompany $M$-matrix
  \beq\label{x52}
  \begin{array}{c}
  \displaystyle{
 {\mathcal M}^{\hbox{\tiny{CM}}}_{ab}=\nu\delta_{ab}
d_a+\nu(1-\delta_{ab})F_{ab}^{z}+\nu\delta_{ab}\,\mathcal F^0\,,\ \
\ F_{ab}^{z}=\p_{q_a}R_{ab}^{z}(q_a-q_b)\,,
 }
 \end{array}
 \eq
 where
   \beq\label{x53}
  \begin{array}{c}
  \displaystyle{
d_a=-\sum\limits_{c:\,c\neq a}^{M}F^0_{ac}\,,\ \ \ \mathcal
F^0=\sum\limits_{b,c:\,b> c}^{M}F^0_{bc}=\sum\limits_{b,c:\,b>
c}^{M}\p_{q_b}r_{bc}(q_{bc})\,.
 }
 \end{array}
 \eq
In the $N=1$ case this Lax pair coincides with the widely known
Krichever's result \cite{Krich1} for ${\rm gl}_M$ Calogero-Moser
model. The last term  $\mF^0$ in (\ref{x52}) enters $\mM$ as a
scalar (it is an identity matrix in $\MatM$ component) in the
auxiliary space $\MatM$. Therefore, it can be moved to the l.h.s. of
the Lax equation. This yields
  \beq\label{q311}
  \begin{array}{c}
  \displaystyle{
\{H^{\hbox{\tiny{CM}}},\mL^{\hbox{\tiny{CM}}}\}+[\nu\mF^0,\mL^{\hbox{\tiny{CM}}}(z)]
=[\mL^{\hbox{\tiny{CM}}}(z),\bar{\mM}^{\hbox{\tiny{CM}}}(z)]\,,
 }
 \end{array}
 \eq
 where
$\bar\mM^{\hbox{\tiny{CM}}}=\mM^{\hbox{\tiny{CM}}}-\nu
1_M\otimes\mF^0$. On the one hand (\ref{q311}) is just a rewritten
classical Lax equation for the spinless Calogero-Moser model. On the
other hand we may treat it as half-quantum Lax equation in a sense
that the dynamics is given by the interacting tops Hamiltonian
(\ref{t81}), where the spin variables are already quantized, while
the positions and momenta remain classical. Indeed, the quantization
of $\mS^{ii}_1$ in fundamental representation of ${\rm GL}_N$ is
given by the permutation operator $P_{1j}$. Plugging it into the
potential of (\ref{t81}) we get the $\mF^0$ term from (\ref{x52})
and (\ref{q311}).

Thus the $R$-matrix valued Lax pairs are multidimensional classical
Lax pairs for the spinless Calogero-Moser models and at the same
time they are quantum Lax pairs for the models of interacting tops
with the spin variables being quantized in the fundamental
representation of ${\rm GL}_N$, i.e. the $\mF^0$ term is the
quantization of the potential $\mV(\mS^{ii},\mS^{jj},q_i-q_j)$
(\ref{t209}).

Let us also mention that there is another class of integrable models
with the Hamiltonian of type (\ref{t203}). These are the Gaudin type
models \cite{Nekr}. The corresponding Lax matrix is of size $M\times
M$. It has simple poles at $n$ points on elliptic curve (or its
degenerations) with the classical spin variables matrices attached
to each point. The number of points is not necessarily equal to $M$.
It is an interesting task to find interrelations between the Gaudin
models and the models of interacting tops through the spectral
duality \cite{MMZZ} based on the rank-size duality transformation.


The classical spinless Calogero-Moser model possesses an equilibrium
position, where $p_i=0$ and $q_i=x_i$ (for example, $x_i=i/M$
\cite{CSasaki}). At this point the term
$\{H^{\hbox{\tiny{CM}}},\mL^{\hbox{\tiny{CM}}}\}$ vanishes from the
l.h.s. of (\ref{q311}), and we are left with the quantum Lax
equation for some long-range (quantum) spin chain. It is an
anisotropic generalization \cite{SeZ} of the
Haldane-Shastry-Inozemtsev type chains. An open question is which
$\mF^0$ provide integrable spin chains? To confirm integrability we
need to construct higher Hamiltonians, which commute with each other
and with $\mF^0(q_i=x_i)$. Taking into account all the above we
guess that the model of interacting tops together with the freezing
trick (the quantum version of the equilibrium position) can be used
to calculate higher spin chain Hamiltonians and to prove their
commutativity. For this purpose we need to construct a quantization
for the model of interacting tops, which is the subject of our next
paper.

Another one intriguing question is to construct relativistic
generalization of the models discussed above. While the classical
models of relativistic interacting tops are expected to be
relatively simple (the block $L^{ij}$ in (\ref{t51}) should be
replaced by $\tr_2(\mS^{\,ij}_2\,R^z_{12}(q_{ij}+\eta)P_{12})$) its
quantum versions and the related long-range spin chain were not
studied yet as well as the corresponding $R$-matrix valued Lax
pairs.


\section{Appendix}

\subsection{A: Definitions and identities}
\def\theequation{A.\arabic{equation}}
\setcounter{equation}{0}

The following set of functions is used in this paper \cite{Weil}.
The first one is the Kronecker function:
  \beq\label{a01}
  \begin{array}{l}
  \displaystyle{
 \phi(\eta,z)=\left\{
   \begin{array}{l}
    1/\eta+1/z\quad - \quad \hbox{rational case}\,,
    \\
    \coth(\eta)+\coth(z) \quad - \quad \hbox{trigonometric case}\,,
    \\
    \frac{\vth'(0)\vth(\eta+z)}{\vth(\eta)\vth(z)} \quad - \quad
    \hbox{elliptic
    case}\,.
   \end{array}
 \right.
 }
 \end{array}
 \eq
 Its elliptic version is given in terms of the odd theta-function
   \beq\label{a02}
  \begin{array}{c}
  \displaystyle{
   \vth(z)=\displaystyle{\sum _{k\in \mathbb Z}} \exp \left ( \pi
\imath \tau (k+\frac{1}{2})^2 +2\pi \imath
(z+\frac{1}{2})(k+\frac{1}{2})\right )
 }
 \end{array}
 \eq
on elliptic curve with moduli $\tau$ (Im$(\tau)>0$). Next are the
first Eisenstein (odd) function and the Weierstrass (even)
$\wp$-function:
  \beq\label{a03}
  \begin{array}{c}
  \displaystyle{
 E_1(z)=\left\{
   \begin{array}{l}
 1/z\,,
\\
   \coth(z)\,,
\\
    \vth'(z)/\vth(z)\,,
   \end{array}
 \right.\hskip12mm  \wp(z)=\left\{
   \begin{array}{l}
 1/z^2\,,
\\
   1/\sinh^2(z)\,,
\\
    -\p_z E_1(z)+\frac{1}{3}\frac{\vth'''(0)}{\vth'(0)}\,.
   \end{array}
 \right.
 }
 \end{array}
 \eq
 We also need the derivatives
 \beq\label{a04}
  \begin{array}{c}
  \displaystyle{
   E_2(z)= -\p_z E_1(z)
 }
 \end{array}
 \eq
and
  \beq\label{a05}
  \begin{array}{l}
  \displaystyle{
 f(z,q)\equiv\p_q\phi(z,q)=\phi(z,q)(E_1(z+q)-E_1(q))\,.
 }
 \end{array}
 \eq
The one (\ref{a04}) is the second Eisenstein function.

The main relation is the Fay trisecant identity:
  \beq\label{a06}
  \begin{array}{c}
  \displaystyle{
\phi(z,q)\phi(w,u)=\phi(z-w,q)\phi(w,q+u)+\phi(w-z,u)\phi(z,q+u).
 }
 \end{array}
 \eq
  The following degenerations of (\ref{a06}) are necessary for the Lax
  equations and $r$-matrix structures:
  \beq\label{a07}
  \begin{array}{c}
  \displaystyle{
 \phi(z,x)f(z,y)-\phi(z,y)f(z,x)=\phi(z,x+y)(\wp(x)-\wp(y))\,,
 }
 \end{array}
 \eq
  \beq\label{a08}
  \begin{array}{c}
  \displaystyle{
 \phi(\eta,z)\phi(\eta,-z)=\wp(\eta)-\wp(z)=E_2(\eta)-E_2(z)\,.
 }
 \end{array}
 \eq
 Also
  \beq\label{a10}
  \begin{array}{c}
  \displaystyle{
 \phi(z,q)\phi(w,q)=\phi(z+w,q)(E_1(z)+E_1(w)+E_1(q)-E_1(z+w+q))=
 }
 \\ \ \\
  \displaystyle{
 =\phi(z+w,q)(E_1(z)+E_1(w))-f(z+w,q)\,.
 }
 \end{array}
 \eq
 The local behavior of the Kronecker function and the first Eisenstein function near its simple pole
 at $z=0$ is as follows:
 \beq\label{a11}
 \begin{array}{c}
  \displaystyle{
 \phi(z,u)=\frac{1}{z}+E_1(u)+\frac{z}{2}\,(E_1^2(u)-\wp(u))+
 O(z^2)\,,
  }
 \end{array}
 \eq
 \beq\label{a12}
 \begin{array}{c}
  \displaystyle{
 E_1(z)=\frac{1}{z}+\frac{z}{3}\,\frac{\vth'''(0)}{\vth'(0)}+O(z^3)\,.
  }
 \end{array}
 \eq
 From (\ref{a11}) and (\ref{a05}) it follows that
   \beq\label{a13}
 \begin{array}{c}
  \displaystyle{
 f(0,u)=-E_2(u)\,.
  }
 \end{array}
 \eq


\subsection{B: Spin ${\rm gl}_M$ Calogero-Moser model}
\def\theequation{B.\arabic{equation}}
\setcounter{equation}{0}
The Lax equations
  \beq\label{a241}
  \begin{array}{c}
  \displaystyle{
{\dot
L}^{\hbox{\tiny{spin}}}(z)=[L^{\hbox{\tiny{spin}}}(z),M^{\hbox{\tiny{spin}}}(z)]
 }
 \end{array}
 \eq
 with the Lax pair
  \beq\label{a24}
  \begin{array}{c}
  \displaystyle{
L^{\hbox{\tiny{spin}}}_{ij}(z)=\delta_{ij}(p_i+S_{ii}E_1(z))+(1-\delta_{ij})S_{ij}\phi(z,q_{ij})\,,
 }
 \end{array}
 \eq
  \beq\label{a36}
  \begin{array}{c}
  \displaystyle{
  M^{\hbox{\tiny{spin}}}(z)_{ij}=(1-\delta_{ij})S_{ij}f(z,q_i-q_j)\,.
 }
 \end{array}
 \eq
 provide (after restriction on the constraints (\ref{t07})) equations of motion
  \beq\label{a33}
  \begin{array}{c}
  \displaystyle{
{\dot q}_i=p_i\,,\quad {\ddot q}_i=\sum\limits_{j\neq
i}^MS_{ij}S_{ji}\wp'(q_i-q_j)\,,
 }
 \end{array}
 \eq
  \beq\label{a34}
  \begin{array}{c}
  \displaystyle{
{\dot S}_{ii}=0\,,\quad {\dot S}_{ij}=\sum\limits_{k\neq i,j}^M
S_{ik}S_{kj}(\wp(q_i-q_k)-\wp(q_j-q_k))\,,\ i\neq j\,.
 }
 \end{array}
 \eq
 The l.h.s. of the Lax equations (\ref{a241}) is generated by  the Hamiltonian (\ref{t06})
  \beq\label{a251}
  \begin{array}{c}
  \displaystyle{
{\dot
L}^{\hbox{\tiny{spin}}}(z)=\{H^{\hbox{\tiny{spin}}},L^{\hbox{\tiny{spin}}}(z)\}
 }
 \end{array}
 \eq
 and the linear Poisson-Lie brackets on ${\rm gl}_M^*$:
  \beq\label{a25}
  \begin{array}{c}
  \displaystyle{
\{S_{ij},S_{kl}\}=-S_{il}\delta_{kj}+S_{kj}\delta_{il}\quad
\hbox{or}\quad \{S_1,S_2\}=[S_2,P_{12}]\,.
 }
 \end{array}
 \eq
 Recall that the Poisson reduction with respect to Cartan action
 (\ref{t08}) is non-trivial. For instance, in the rank 1 case
 (\ref{t09}) such reduction leads to the spinless model (\ref{t10}).
 Explicit expression of the reduced Poisson structure depends on a
 choice of gauge fixation conditions. The equations of motion
 (\ref{a33})-(\ref{a34}) are not the reduced. They are obtained by a simple restriction. To get the final
 equations one should perform the Dirac reduction and evaluate the
 Dirac terms.

The classical $r$-matrix structure is as follows:
  \beq\label{a28}
  \begin{array}{c}
  \displaystyle{
\{L^{\hbox{\tiny{spin}}}_1(z),L^{\hbox{\tiny{spin}}}_2(w)\}=[L^{\hbox{\tiny{spin}}}_1(z),r^{\hbox{\tiny{spin}}}_{12}(z,w)]
-[L^{\hbox{\tiny{spin}}}_2(w),r^{\hbox{\tiny{spin}}}_{21}(w,z)]-
 }
 \\ \ \\
  \displaystyle{
 -\sum\limits_{i\neq j} E_{ij}\otimes E_{ji} (S_{ii}-S_{jj})
f(z-w,q_{ij})
 }
 \end{array}
 \eq
with
   \beq\label{a21}
  \begin{array}{c}
  \displaystyle{
 r^{\hbox{\tiny{spin}}}_{12}(z,w)=E_1(z-w)\sum\limits_{i=1}^ME_{ii}\otimes
 E_{ii}
+\sum\limits_{i\neq j}^M\phi(z-w,q_{ij})\,E_{ij}\otimes E_{ji}\,.
 }
 \end{array}
 \eq
Here the linear Poisson brackets (\ref{a25}) are assumed as well.
The Dirac reduction is not yet performed. However, we can see that
the restriction on the constraints (\ref{t07}) kills the last term
in (\ref{a28}), and we are left with the standard linear classical
$r$-matrix structure. It is enough for Poisson commutativity
   \beq\label{a29}
  \begin{array}{c}
  \displaystyle{
  \{\tr(L^k(z)),\tr(L^n(w))\}=0\,,\quad\forall\, k,n\in\mZ_+\,,\
  z,w\in\mC
 }
 \end{array}
 \eq
necessary for the Liouville integrability. The proof of (\ref{a28})
is direct. It is based on the identities (\ref{a06})-(\ref{a10}).
Let us write down a few examples of verification of (\ref{a28}):

\underline{the tensor component $E_{ij}\otimes E_{jk}$ ($ i\neq j\,,
 j\neq k\,, k\neq i$):}

\noindent l.h.s. of (\ref{a28}):
    \beq\label{a51}
   \begin{array}{c}
  \displaystyle{
  \{L^{\hbox{\tiny{spin}}}_{ i j}(z),L^{\hbox{\tiny{spin}}}_{ j k}(w)\}=
  \{S_{ i j},S_{ j k}\}\phi(z,q_{ i j})\phi(w,q_{ j k})=
  -S_{ i k}\,\phi(z,q_{ i j})\phi(w,q_{ j k})\,.
 }
 \end{array}
 \eq
\noindent r.h.s. of (\ref{a28}):
    \beq\label{a52}
   \begin{array}{c}
  \displaystyle{
  S_{ i k}\phi(z,q_{ i k})\phi(z-w,q_{ k j})+S_{ i k}\phi(w,q_{ i k})\phi(w-z,q_{ j i})\,.
 }
 \end{array}
 \eq
 Expressions (\ref{a51}) and (\ref{a52}) coincide due to
 (\ref{a06}).

\underline{the tensor component $E_{ i i}\otimes E_{ i j}$ ($ i\neq
j$):}

\noindent l.h.s. of (\ref{a28}):
    \beq\label{a53}
   \begin{array}{c}
  \displaystyle{
  \{L^{\hbox{\tiny{spin}}}_{ i i}(z),L^{\hbox{\tiny{spin}}}_{ i j}(w)\}=
  \{p_{ i},\phi(w,q_{ i j})\}S_{ i j}+
  \{S_{ i i},S_{ i j}\}E_1(z)\phi(w,q_{ i j})=
 }
 \\ \ \\
  \displaystyle{
  =S_{ i j}f(w,q_{ i j})-
  S_{ i j}E_1(z)\phi(w,q_{ i j})\,.
 }
 \end{array}
 \eq
\noindent r.h.s. of (\ref{a28}):
    \beq\label{a54}
   \begin{array}{c}
  \displaystyle{
  S_{ i j}\phi(z,q_{ i j})\phi(z-w,q_{ j i})+S_{ i j}E_1(w-z)\phi(w,q_{ i j})\,.
 }
 \end{array}
 \eq
 Expressions (\ref{a53}) and (\ref{a54}) coincide due to
 (\ref{a10}).

 \underline{the tensor component $E_{ i j}\otimes E_{ j i}$
($ i\neq j$):}

\noindent l.h.s. of (\ref{a28}):
    \beq\label{a55}
   \begin{array}{c}
  \displaystyle{
  \{L^{\hbox{\tiny{spin}}}_{ i j}(z),L^{\hbox{\tiny{spin}}}_{ j i}(w)\}=
 }
 \\ \ \\
  \displaystyle{
  =\{S_{ i j},S_{ j i}\}\phi(z,q_{ i j})\phi(w,q_{ j i})=(S_{ i i}-S_{ j j})\phi(z,q_{ i j})\phi(-w,q_{ i j})\,.
 }
 \end{array}
 \eq
The last term from the r.h.s. of (\ref{a28}) contributes in this
component. The r.h.s. of (\ref{a28}):
    \beq\label{a56}
   \begin{array}{c}
  \displaystyle{
  (p_ i+S_{ i i}E_1(z)-p_ j-S_{ j j}E_1(z))\phi(z-w,q_{ i j})-
 }
 \\ \ \\
  \displaystyle{
  -(p_ j+S_{ j j}E_1(w)-p_ i-S_{ i i}E_1(w))\phi(w-z,q_{ j i})-(S_{ i i}-S_{ j j})f(z-w,q_{ i j})=
 }
  \\ \ \\
  \displaystyle{
  =(S_{ i i}-S_{ j j})\Big((E_1(z)-E_1(w))\phi(z-w,q_{ i j})-f(z-w,q_{ i j})\Big)\,.
 }
 \end{array}
 \eq
 Expressions (\ref{a55}) and (\ref{a56}) coincide due to
 (\ref{a10}).

\subsection{C: Integrable ${\rm gl}_N$ tops}
\def\theequation{C.\arabic{equation}}
\setcounter{equation}{0}

It was shown in \cite{LOZ16} (see also \cite{LOZ8}) that the Lax
equations
  \beq\label{c405}
  {\dot L}(z,S)=[L(z,S),M(z,S)]
  \eq
are equivalent to equations
  \beq\label{c406}
\dot{S}=[S,J(S)]
  \eq
 for the Lax pair
  \beq\label{c409}
  \begin{array}{c}
     \displaystyle{
 L(z,S)=\tr_2(r_{12}(z)S_2)\,,\qquad
  M(z,S)=\tr_2(m_{12}(z)S_2)\,,\quad S_2=1_N\otimes S
 }
  \end{array}
  \eq
and
  \beq\label{c410}
  \begin{array}{c}
     \displaystyle{
 J(S)=\tr_2(m_{12}(0)S_2)\,.
 }
  \end{array}
  \eq
 constructed by means of the coefficients of the (classical
 limit) expansion (\ref{t004}) for an $R$-matrix satisfying the
 associative Yang-Baxter equation (\ref{t003}) and the properties
 (\ref{t300})-(\ref{t302}). The answer (\ref{c409}) can be written
 more explicitly. For
  \beq\label{c411}
  \begin{array}{c}
     \displaystyle{
 r_{12}(z)=\sum\limits_{i,j,k,l=1}^N r_{ijkl}(z)\,
 e_{ij}\otimes e_{kl}
 }
  \end{array}
  \eq
(\ref{c409}) means
  \beq\label{c412}
  \begin{array}{c}
     \displaystyle{
 L(z,S)=\sum\limits_{i,j,k,l=1}^N r_{ijkl}(z)S_{lk}\,e_{ij}
 }
  \end{array}
  \eq
 since $\tr(e_{kl}S)=S_{lk}$.

 Let us briefly describe how these formulae reproduce the elliptic
 top from \cite{LOZ}. In the elliptic case we need special matrix
 basis in $\Mat$. Consider the matrices
 \beq\label{e903}
 \begin{array}{c}
  \displaystyle{
Q_{kl}=\delta_{kl}\exp\left(\frac{2\pi
 \imath}{N}k\right)\,,\ \ \ \Lambda_{kl}=\delta_{k-l+1=0\,{\hbox{\tiny{mod}}}
 N}\,,\quad Q^N=\Lambda^N=1_{N}\,.
 }
 \end{array}
 \eq
 Then the basis in $\Mat$ is given by the following set:
 \beq\label{e904}
 \begin{array}{c}
  \displaystyle{
 T_a=T_{a_1 a_2}=\exp\left(\frac{\pi\imath}{N}\,a_1
 a_2\right)Q^{a_1}\Lambda^{a_2}\,,\quad
 a=(a_1,a_2)\in\mZ_N\times\mZ_N\,.
 }
 \end{array}
 \eq
 Since
 \beq\label{e9041}
 \begin{array}{c}
  \displaystyle{
 \exp\left(\frac{2\pi\imath}{N}\,a_1
 a_2\right)Q^{a_1}\Lambda^{a_2}=\Lambda^{a_2}Q^{a_1}
 }
 \end{array}
 \eq
we have
  \beq\label{e905}
 \begin{array}{c}
  \displaystyle{
T_\al T_\be=\kappa_{\al,\be} T_{\al+\be}\,,\ \ \
\kappa_{\al,\be}=\exp\left(\frac{\pi \imath}{N}(\be_1
\al_2-\be_2\al_1)\right)\,,
 }
 \end{array}
 \eq
 where $\al+\be=(\al_1+\be_1,\al_2+\be_2)$. The non-degenerate
 pairing is given by the matrix trace:
  \beq\label{e906}
 \begin{array}{c}
  \displaystyle{
\tr(T_\al T_\be)=N\delta_{\al+\be}\,,\quad T_0=1_N\,.
 }
 \end{array}
 \eq
 Define the set of functions numerated by
$a=(a_1,a_2)\in\mZ_N\times\mZ_N$:
 \beq\label{q913}
 \begin{array}{c}
  \displaystyle{
 \vf_a(z,\om_a+u)=\exp(2\pi\imath\frac{a_2}{N}\,z)\,\phi(z,\om_a+u)\,,\quad
 \om_a=\frac{a_1+a_2\tau}{N}
 }
 \end{array}
 \eq
 and introduce notation
 \beq\label{q9131}
 \begin{array}{c}
  \displaystyle{
 f_a(z,\om_a+u)=\exp(2\pi\imath\frac{a_2}{N}\,z)\,f(z,\om_a+u)\,.
 }
 \end{array}
 \eq
 The Baxter-Belavin $R$-matrix \cite{Belavin} satisfying all required properties including the
 Fourier symmetry (\ref{t303}) is of the form:
 \beq\label{c61}
 \begin{array}{c}
  \displaystyle{
 R^{BB}_{12}(\hbar,z)=
 \sum\limits_{a\in\,{\mathbb Z}_N\times{\mathbb Z}_N} \vf_a(z,\hbar+\om_a)\, T_a\otimes
 T_{-a}\in \Mat^{\otimes 2}\,.
 }
 \end{array}
 \eq
This $R$-matrix satisfies required properties but with different
normalizations. For example, the Fourier symmetry has form
$R^{BB}_{12}(\hbar,z)P_{12}=R^{BB}_{12}(z/N,N\hbar)$ (see the
Fourier transformation formulae in \cite{Z2}). To fulfill all
requirements including the normalization (\ref{t302}) we consider
 \beq\label{c62}
 \begin{array}{c}
  \displaystyle{
 R^\hbar_{12}(z)=R^{BB}_{12}(\hbar/N,z)=\frac{1}{N}
 \sum\limits_{a\in\,{\mathbb Z}_N\times{\mathbb Z}_N} \vf_a(z,\frac{\hbar}{N}+\om_a)\, T_a\otimes
 T_{-a}\in \Mat^{\otimes 2}\,.
 }
 \end{array}
 \eq
 The corresponding classical $r$-matrix is as follows
 \beq\label{c63}
 \begin{array}{c}
  \displaystyle{
 r_{12}(z)=\frac{1}{N}\,E_1(z)\,1_N\otimes1_N+\frac{1}{N}
 \sum\limits_{a\neq(0,0)} \vf_a(z,\om_a)\, T_a\otimes
 T_{-a}\in \Mat^{\otimes 2}\,,
 }
 \end{array}
 \eq
and
 \beq\label{c64}
 \begin{array}{c}
  \displaystyle{
 m_{12}(z)=\frac{E_1^2(z)-\wp(z)}{2N^2}\,1_N\otimes 1_N+\frac{1}{N^2}
 \sum\limits_{a\neq(0,0)} f_a(z,\om_a)\, T_a\otimes
 T_{-a}\in \Mat^{\otimes 2}\,.
 }
 \end{array}
 \eq
 Then the formulae for the Lax pair (\ref{c409}) reproduce the Lax
 pair of the elliptic top. It is contained in the Lax pair
 (\ref{t20})-(\ref{t22}) as a diagonal $N\times N$ block.

 The derivative of the classical $r$-matrix is obtained through
 (\ref{a05}):
 \beq\label{c65}
 \begin{array}{c}
  \displaystyle{
 F^0_{12}(z)=\p_zr_{12}(z)=-\frac{1}{N}\,E_2(z)\,1_N\otimes1_N+
 }
 \\ \ \\
  \displaystyle{
 +\frac{1}{N}\sum\limits_{a\neq(0,0)} \vf_a(z,\om_a)(E_1(z+\om_a)-E_1(z)+2\pi\imath\p_\tau\om_\al)\, T_a\otimes
 T_{-a}\,.
 }
 \end{array}
 \eq
 The Fourier symmetry $R^\hbar_{12}(z)=R^z_{12}(\hbar)P_{12}$ for the $R$-matrix (\ref{c62}) is
 based on the following set of identities for the functions (\ref{q913}):
  \beq\label{c66}
 \begin{array}{c}
  \displaystyle{
\frac{1}{{ N}}\sum\limits_{\al} \kappa_{\al,\ga}^2\, \vf_\al({
N}\hbar,\om_\al+\frac{z}{{ N}})=\vf_\ga(z,\om_\ga+\hbar)\,,\quad
\forall\ga\in\mZ_N\times\mZ_N\,.
  }
 \end{array}
 \eq
 Be degeneration of the latter identities on can deduce (see
 \cite{Z2}):
  \beq\label{c67}
 \begin{array}{c}
  \displaystyle{
 \sum\limits_{\al} E_2(\om_\al+q)={ N}^2E_2({ N}q)
  }
 \end{array}
 \eq
 and for $\ga\neq 0$
  \beq\label{c68}
 \begin{array}{c}
  \displaystyle{
 \sum\limits_{\al} \kappa_{\al,\ga}^2 E_2(\om_\al+q)=-
 { N}^2\vf_\ga({ N}q,\om_\ga)
 (E_1({ N}q+\om_\ga)-E_1({ N}q)+2\pi\imath\p_\tau\om_\ga)\,.
  }
 \end{array}
 \eq
Conversely, 
  \beq\label{c69}
 \begin{array}{c}
  \displaystyle{
 -E_2(q)+\sum\limits_{\al} \kappa_{\al,\ga}^2 \vf_\al(q,\om_\al)
 (E_1(q+\om_\al)-E_1(q)+2\pi\imath\p_\tau\om_\al)=-
 E_2(\om_\ga+\frac{q}{N})\,.
  }
 \end{array}
 \eq



\paragraph{Acknowledgments.} The work
was partially supported by RFBR grants 18-01-00926 (all authors) and
19-51-18006 Bolg$_a$ (I. Sechin). The research of A. Zotov was also
supported in part by the HSE University Basic Research Program,
Russian Academic Excellence Project '5-100' and by the Young Russian
Mathematics award. A. Zotov is also grateful to Sasha Alexandrov for
hospitality at the 2-nd Workshop on integrable systems and
applications (May, 2019) in IBS Center for Geometry and Physics
(Pohang, Korea), where a part of the results was presented.

\begin{small}
 
\end{small}

\end{document}